\documentclass[aps,prl,twocolumn,tightenlines,amsmath,amssymb,superscriptaddress]{revtex4-1}
\usepackage{graphicx}
\usepackage{amsmath}
\usepackage{amssymb}
\usepackage{epstopdf}
\usepackage{lineno}
\usepackage{color}
\usepackage{lipsum}
\usepackage{ulem}
\usepackage[dvipsnames]{xcolor}
\usepackage[colorlinks=true]{hyperref} 
\usepackage[sort&compress]{natbib}
\usepackage{braket}
\usepackage{lineno}

\newcommand*{\rom}[1]{\expandafter\@slowromancap\romannumeral #1@}

\begin{document}

\title{Experimental analysis of energy transfers between a quantum emitter and light fields}

\author
{{I. Maillette de Buy Wenniger$^{1}$, S. E. Thomas$^1$, M. Maffei$^2$, S. C. Wein$^{2,3}$,\\ 
M. Pont$^1$, 
{N. Belabas$^1$, 
S. Prasad$^2$, }
A. Harouri$^1$, A. Lema\^itre$^1$, I. Sagnes$^1$, N. Somaschi$^4$,\\ 
A. Auff\`eves$^{\ast 5,6}$, P. Senellart$^\ast$}}
	
\affiliation{Centre for Nanosciences and Nanotechnology, CNRS, Universit\'e Paris-Saclay, UMR 9001,
10 Boulevard Thomas Gobert, 91120 Palaiseau, France\\
    $^2$Universit\'{e} Grenoble Alpes, CNRS, Grenoble INP, Institut N\'eel, 38000 Grenoble, France\\
    $^3$Institute for Quantum Science and Technology and Department of Physics and Astronomy,
University of Calgary, Calgary, Alberta, Canada T2N 1N4 \\
    $^4$Quandela SAS, 10 Boulevard Thomas Gobert, 91120 Palaiseau, France\\
$^5$ MajuLab, CNRS-UCA-SU-NUS-NTU International Joint Research Laboratory\\
$^6$Centre for Quantum Technologies, National University of Singapore, 117543 Singapore, Singapore}

\begin{abstract}
Energy transfer between quantum systems can either be achieved through an effective unitary interaction or through the generation of entanglement. This observation defines two types of energy exchange: unitary and correlation energy. Here we propose and implement experimental protocols to access these energy transfers in interactions between a quantum emitter and light fields. Upon spontaneous emission, we measure the unitary energy transfer from the emitter to the optical field and show that it never exceeds half of the total energy and is reduced when introducing decoherence. We then study the interference of the emitted field and a laser field at a beam splitter and show that the energy transfers quantitatively depend on the quantum purity of the emitted field.
\end{abstract}

\maketitle

The study of energy exchanges between closed, isolated quantum systems is highly relevant to quantum technologies, be it to study quantum batteries~\cite{Quantum_battery,Quantacell,Polini2019} or the fundamental energetic cost of  quantum information processing~\cite{Stevens2022,GB,Ozawa}. These studies are important to understand the energetic costs not only at the fundamental level, but also at the macroscopic level~\cite{Alexia}. In this context, understanding and probing the nature of energy exchanges is essential. 
In the canonical case of a bipartite system, two different types of energy transfers can be identified. We  consider two  quantum systems $A$ and $B$ with the energy initially in $A$ and subsequently transferred to $B$ such that at the end of the interaction we have $-\Delta {\cal E}_A = \Delta {\cal E}_B$. After the interaction, we can define the unitary energy ${E}^{A,B}_\mathrm{unit}$ as the energy transferred from $A$ to $B$ through an effective unitary interaction from the point of view of $B$. This unitary energy is limited by quantum correlations that build up during the interaction between the two systems. The remaining energy, dubbed correlation energy ${E}^{A,B}_\mathrm{corr}= \Delta {\cal E}_B-{E}^{A,B}_\mathrm{unit}$, signals the presence of correlations or entanglement during the interaction. In the limit where $A$ becomes classical, correlations disappear and $B$ only receives unitary energy. From a quantum thermodynamics perspective, this corresponds to the unambiguous case where a closed quantum system $B$ is classically driven, such that it only receives work \cite{Binder2018}. As such, unitary energy is naturally associated with a notion of work for closed quantum systems~\cite{Alipour2016,Hossein-Nejad2015,Weimer2008,Schroder2018}.

Theoretically, this framework has been applied to the the case of a qubit coupled to a reservoir of electromagnetic modes~\cite{Ciccarello2021,fan2010,landi2019}, a situation relevant for quantum light generation and qubit manipulation~\cite{Andolina2018, Ferraro2018, Monsel2020, Maffei2021}. As an example, when a coherent and intense classical field drives a qubit, both systems remain non-entangled and only exchange unitary energy: the qubit purity is unaltered, an ideal regime to implement single qubit gates~\cite{Mikko2017,Monsel2020}. The case of spontaneous emission, where a quantum emitter releases its energy into a mode of the electromagnetic field has theoretically been studied. It was predicted that the fraction of unitary energy transferred is proportional to the initial quantum coherence between the ground and excited state of the qubit and represents at most 50~\% of the total energy transferred~\cite{Monsel2020}. 

\begin{figure*}
\includegraphics[width=\linewidth]{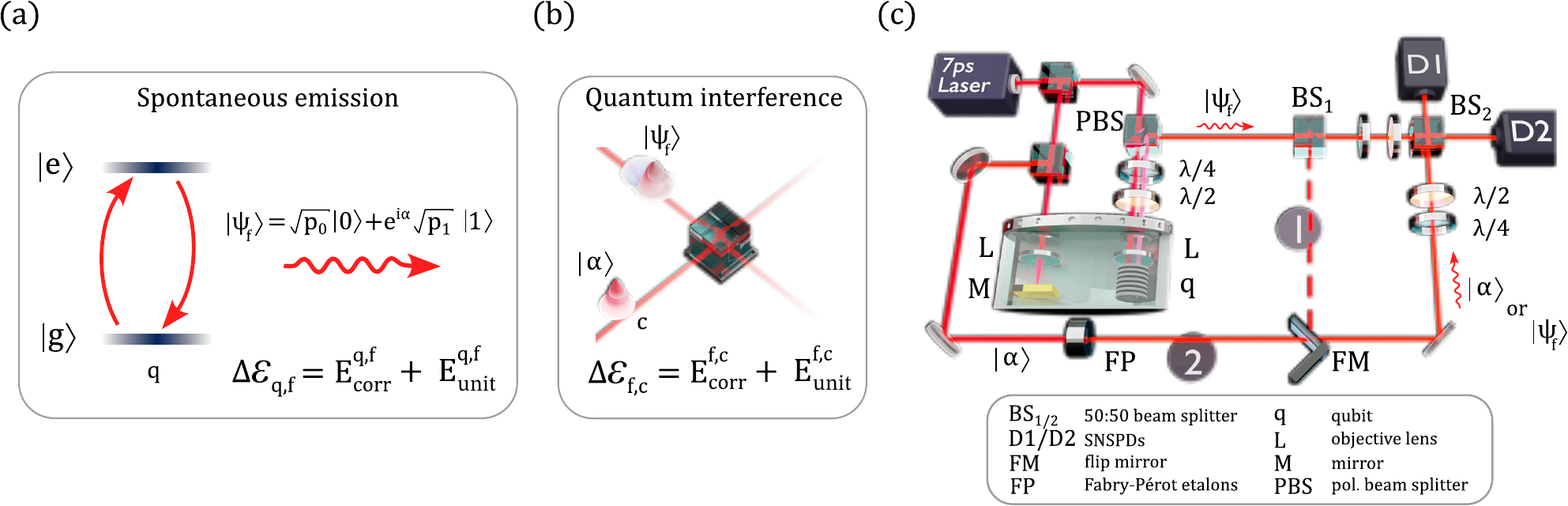}
\caption{\label{fig:1} \textbf{Measuring energetic transfers.} (a) Sketch of the first scenario studied: energetic transfers in spontaneous emission, from a quantum emitter ``q'' to the vacuum of the electromagnetic field ``f''. (b) The second scenario studied: energetic transfers from the spontaneously emitted field to a classical field ``c'' at a beam splitter. (c) Experimental implementation: a resonant pulsed laser excites the QD-cavity device. The emitted photonic field is filtered from the laser drive in cross-polarization at a polarizing beam splitter (PBS). In configuration (1) with an additional flip mirror (FM), the photonic field path is split into two paths at a beam splitter (BS$_1$) and one arm is delayed such that two subsequently emitted photonic fields interfere at a 50:50 beam splitter (BS$_2$). In configuration (2) the classical field is derived from the same drive laser, reflected off a mirror (M) in the same cryostation as the QD device ``q'' to minimize the impact of mechanical vibrations and shaped using a Fabry-P\'erot etalon (FP). Removing the flip mirror, the quantum and classical fields interfere at BS$_2$. For both steps, the output intensities are recorded using two superconducting nanowire single-photon detectors D1 and D2.}
\end{figure*}

Experimentally exploring these situations is very challenging: it not only requires the ability to accurately access correlated (i.e. respecting global energy conservation) energy changes within two coupled quantum systems, but also to trace back how these systems exchanged energy. So far experiments have focused on energy changes experienced by only one (open) quantum system, e.g. polarized photons \cite{cimini2020}, ions \cite{VonLindenfels2019}, spins in MNR \cite{Peterson2019}, or on average energy exchanges between two systems, like a superconducting qubit and a field \cite{Cottet2017,Stevens2022}. Here, we take a leap forward, by proposing and implementing experimental protocols to access the unitary and correlation energies {exchanged in a closed bipartite system} {in two scenarios (Fig.~\ref{fig:1}(a,b)): the spontaneous emission of a quantum emitter, and interference between the emitted field and a classical field}. The emitter is a single semiconductor quantum dot inserted in an optical microcavity. We experimentally access the fraction of the unitary energy emitted during spontaneous emission by performing self-homodyne measurements with the emitted light field. At low temperature, we measure a fraction very close to the $50\,\%$ theoretical limit. By increasing the emitter temperature, we study the impact of pure dephasing and observe a reduction in the fraction of unitary energy transfer evidencing loss of coherence induced by the environment of the emitter. In the second scenario, we probe energetic exchanges between the spontaneously emitted field and a coherent field by interfering them on a beam splitter. The energy transferred is shown to be limited by the purity of the quantum field and the relative coherence of both fields.

The energy exchanges between a qubit and a field during spontaneous emission are theoretically described in Ref.~\cite{Monsel2020}. A qubit (q) resonantly excited by a laser drive is brought into the pure quantum superposition state $\ket{\Psi_\mathrm{q}}=\cos(\theta/2)\ket{g}+\sin(\theta/2)\mathrm{e}^{i\alpha}\ket{e}$, with $\ket{g}$ and $\ket{e}$  ground and excited states separated by an energy $\hbar \omega_0$, and $\theta, \alpha \in [0,\pi]$ are the pulse area and the classical phase of the driving laser, respectively. The total initial energy brought to the emitter is $\mathcal{E}_\mathrm{q} = \hbar \omega_0  \sin^2 (\theta/2)$. At the end of the spontaneous emission process, this energy is entirely transferred to the emitted field (f), $\Delta \mathcal{E}_\mathrm{q,f} =\mathcal{E}_\mathrm{q}$. In the absence of decoherence, the emitted field state is pure and reads $\ket{\Psi_\mathrm{f}}=\cos(\theta/2)\ket{0}+\sin(\theta/2)\mathrm{e}^{i\alpha}\ket{1}$, where $\ket{0}$  and $\ket{1}$ are the photon-number states in the mode of the electromagnetic field with populations $p_0 =\cos^2(\theta/2)$ and $p_1 =\sin^2(\theta/2)$, see Fig.~\ref{fig:1}(a). The unitary energy provided by the qubit to the field corresponds to the coherent part of the emitted field energy~\cite{Monsel2020,Maffei2021,Cottet2017} {and} reads $E^\mathrm{q,f}_\mathrm{unit}= \hbar \omega_0  s^2 $, where  $s$ is the coherence of the qubit before spontaneous emission: $s=\cos(\theta/2) \sin(\theta/2)$.  The fraction of unitary energy {transferred} is thus maximal at $\theta=\pi/2$ and amounts to $50\,\%$ of the initial energy. Conversely, the correlation energy corresponds to the incoherent component of the emitted field and corresponds to $E^\mathrm{q,f}_\mathrm{corr}=\Delta \mathcal{E}_\mathrm{q,f}-E^\mathrm{q,f}_\mathrm{unit}$. It is  maximal for $\theta = \pi$ where all the energy is transferred in the form of correlation energy.

\begin{figure}[t]
		\includegraphics[width=0.9\linewidth]{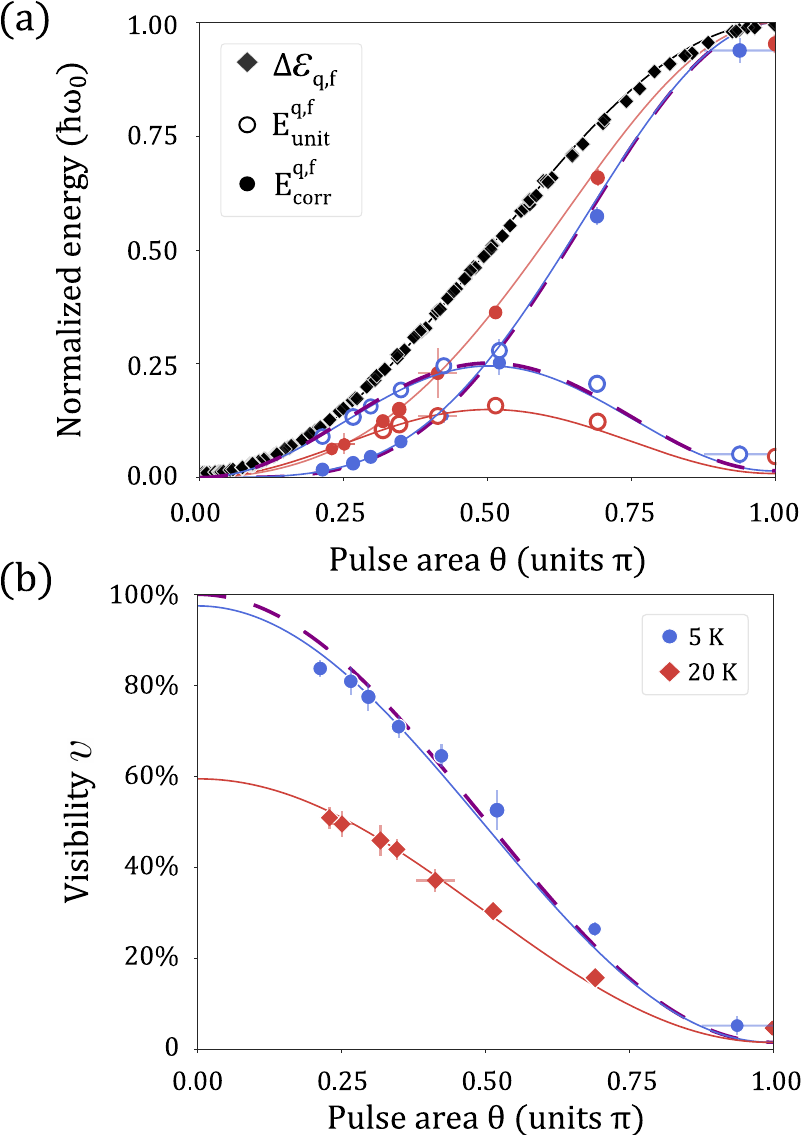}
\caption{\textbf{Energy transfer during spontaneous emission.}  (a) Total energy transferred from the qubit ``q" to the vacuum of the electromagnetic field ``f" $\Delta\mathcal{E}_\mathrm{q,f}$ (diamonds), unitary energy  $E^\mathrm{q,f}_\mathrm{unit}$ (open circles) and correlation energy $E^\mathrm{q,f}_\mathrm{corr}$ (filled circles). (b) Measured visibility $v$ of the photonic field self-homodyne interference as a function of the pulse area $\theta$. For both plots, blue (red) symbols corresponds to measurements at 5~K (20~K) respectively.  The purple dashed lines correspond to theoretical expectations without any decoherence. The red and blue lines correspond to the theoretical expectations considering the introduction of pure dephasing, and with small corrections accounting for imperfect laser rejection.}\label{fig:2} 
\end{figure}

We experimentally probe these energy processes with an InGaAs quantum dot (QD) coupled to a micropillar cavity~\cite{giesz2016, Somaschi2016}. At a temperature of 5~K, these artificial atoms have been shown to be close to the text book atom-cavity system with negligible influence of the solid-state environment. This is attested by the generation of single photons with near-unity indistinguishability with similar devices~\cite{Somaschi2016} and the unique demonstration that they can emit light wave packets in a quantum superposition of 0 and 1 photon~\cite{Loredo2019}. 
{The QD in a cryostation is resonantly driven by  7~ps laser pulses at 925~nm derived from a Ti:Sapphire laser operating at a 81~MHz repetition rate (see Fig.~\ref{fig:1}(c)). The emitted photonic field is separated from the laser drive using a cross-polarization configuration. Black symbols in Fig.~\ref{fig:2}(a) correspond to the normalized intensity $\mu_\text{f}$ of the emitted field as a function of the pulse area, $\theta$, of the driving laser.  We observe the onset of Rabi oscillations attesting the coherent control over the qubit, i.e. the ability to generate arbitrary quantum superpositions of the qubit ground and excited state. We assume a near-unity occupation of the qubit excited state at the power $ P_\pi$  corresponding to the highest intensity so that the pulse area is $\theta=2\arcsin(\sqrt{P/P_\pi})$. {This hypothesis is validated by our measurement that allows to actually measure the excited state probability as we show in Supplementary Fig.~\ref{fig:rabipop}}. This experimental curve then corresponds to the normalized energy of the emitted field  $\Delta{\cal E}_\mathrm{q,f} /(\hbar \omega_0)={\mu_\text{f}}  = \sin^2 (\theta/2)$ (black line). }

To experimentally measure the unitary energy transferred during spontaneous emission, we perform self-homodyne measurements with the emitted field. In the absence of decoherence, we show in the supplementary information that the resulting visibility of interference, $v$, directly corresponds to the fraction of unitary energy {-- i.e. efficiency of unitary energy transfer --} $v=E^\mathrm{q,f}_\mathrm{unit}/\Delta \mathcal{E}_\mathrm{q,f}=\cos^2(\theta/2)$. As shown in configuration 1 of Fig.~\ref{fig:1}(c), we temporally overlap two successively emitted fields separated in time by 12.5~ns at beam splitter BS$_2$. With two single-photon detectors, we record the single counts that show anticorrelated intensities as a function of the relative phase between the two interfering fields~\cite{Loredo2019}. This allows us to deduce the interference visibility $v$.

Fig.~\ref{fig:2}(b) shows the measured visibility of the interference fringes $v$ as a function of  $\theta$ (see Supplementary). The experimentally measured {visibility} decreases continuously as $\theta$ increases, as theoretically expected.  Combining this visibility measurement with the measurement of the total energy given to the field $\Delta \mathcal{E}_\mathrm{q,f}$, we deduce the amount of energy stemming from unitary interaction between the qubit and the field: {$E^\mathrm{q,f}_\mathrm{unit} = \Delta \mathcal{E}_\mathrm{q,f} \times v $}. Fig.~\ref{fig:2}(a) shows the experimental normalized values of $E^\mathrm{q,f}_\mathrm{unit} $ as a function of $\theta$ (open symbols) as well as the corresponding correlation energy $E^\mathrm{q,f}_\mathrm{corr}=\Delta \mathcal{E}_\mathrm{q,f} \times (1- v) $ (filled symbols) and the corresponding theoretical predictions {without decoherence} (purple dashed lines). 
At low $\theta$, {our observations at 5~K are very close to the theoretical predictions in the absence of decoherence that }most of the energy transferred from the qubit to the electromagnetic field corresponds to unitary energy~\cite{Monsel2020}. Such behavior can be understood considering that, in the low excitation regime, the radiated field stems from the qubit dipole: no light-matter entanglement takes place, and the emitted field is remarkably close to a coherent field {truncated for photon-number components above 1.} Conversely, light-matter entanglement emerges  when qubit population is created for increasing $\theta$. {Although at the end of the spontaneous emission process the qubit and field are in a separable state, quantum correlations taking place during spontaneous emission} reduce the amount of unitary energy eventually transferred to the electromagnetic field, reaching the situation where all the energy is transferred in the form of correlation energy for $\theta=\pi$. {In this situation, the emitted field contains no coherent component and gets completely squeezed. }Remarkably, this is when the qubit acts as a deterministic single-photon source -- a key device for discrete variable optical quantum technologies. {The unitary energy transfer is maximal for $\theta=\pi/2$, with an equipartition of unitary and correlation energy.} This corresponds to a maximal initial coherence in the qubit state.

 {We now explore the effect of decoherence by increasing the QD temperature. At 20~K, the Rabi oscillations are mostly unchanged compared to 5~K (see Suppl.) showing that although  the QD is coupled to the phonon bath, it does not receive energy from it. It is mostly subject to phonon-induced pure dephasing  with the phonon sideband strongly suppressed by the Purcell effect of the micropillar cavity in which the QD is embedded~\cite{Grange2017}. The same theoretical framework still holds (see Suppl.): the unitary energy corresponds to the coherent part of the emitted light field that is }now described by the density matrix $ \hat{\rho}_\text{f} = \cos^2(\theta/2) \hat{\rho}_0 + \sin^2(\theta/2) \hat{\rho}_1 + \cos(\theta/2)\sin(\theta/2) (\hat{\rho}_{01} +  \hat{\rho}_{10})$ where the subscripts 0 and 1 stand for the vacuum or one-photon part of the field, respectively.  $M_\mathrm{s} = \mathrm{Tr}[\hat{\rho}_1^2]$ is the single-photon indistinguishability or purity of the single-photon component in the temporal domain~\cite{santori2002}. The reduction of quantum coherence between the vacuum and the one-photon component is captured by ${\cal C} = \mathrm{Tr}[\hat{\rho}_{01}\hat{\rho}_{10}]$. The unitary energy provided by the qubit to the field now reads:  $E^\mathrm{q,f}_\mathrm{unit}= \hbar \omega_0 \mathcal{C}\cos^2{(\theta/2)}\sin^2{(\theta/2)}$.  {In the case of pure dephasing, we further show that ${\cal C}={M_\mathrm{s}}$ (see Suppl.).}
 
 \begin{figure*}
		\includegraphics[width=\linewidth]{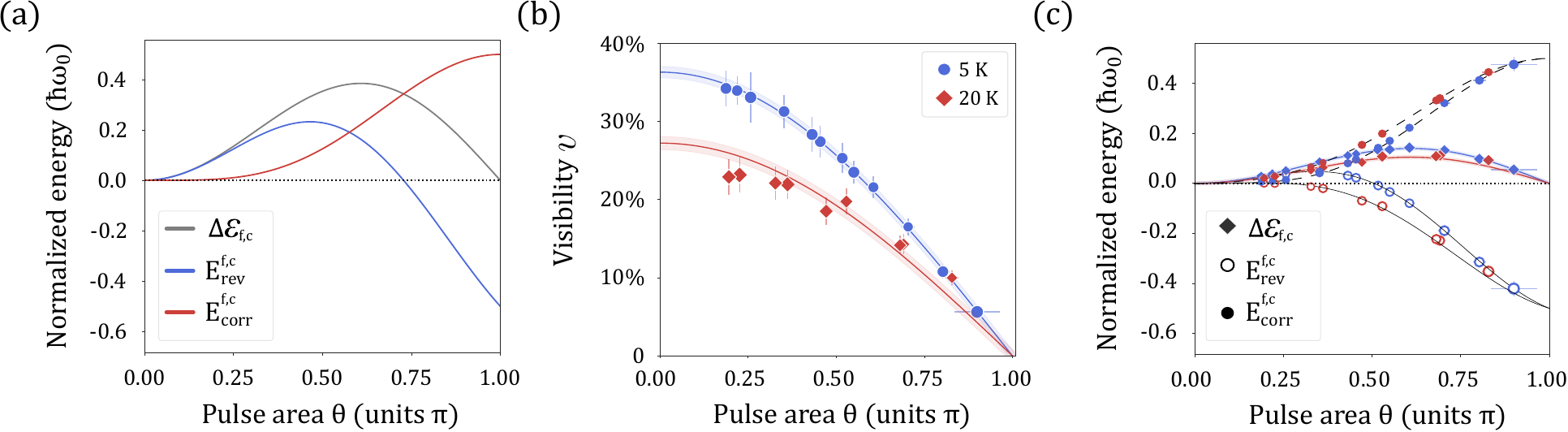}
\caption{ \textbf{Energy exchanges in a quantum interference.}  (a) The theoretical total energy, unitary and correlation energy transfers from the photonic field ``f'' to the classical field ``c'' in the ideal case. (b) Measured visibility $v$ as a function of $\theta$ when interfering the photonic field with the classical field (blue 5~K, red 20~K ). The solid lines are a theoretical fit to the data using $v = \cos(\theta/2)\mathcal{C}_\text{f,c}$, and the shaded regions represent $2\sigma$ uncertainty in the fit. (c) Measured total energy transferred from the quantum field to the classical field, $\Delta\mathcal{E}_\mathrm{f,c}$ as a function of $\theta$, the correlation energy $E^\mathrm{f,c}_\mathrm{corr}$ deduced from Fig.~\ref{fig:2}(a), and the unitary energy transferred $E^\mathrm{f,c}_\mathrm{unit}$ (blue 5~K, red 20~K). Lines correspond to experimental data fit with deduced parameters $\mathcal{C}_{\text{f,c}}$ and $\mathcal{C}$ at each temperature.}\label{fig:3}
	\end{figure*}
 
 Fig.~\ref{fig:2}(b) shows the visibility of the self-homodyne measurement at 20~K  evidencing a {unitary energy} transfer between the QD and the optical field. However, the efficiency is reduced compared to 5~K. We fit both data sets and find $\mathcal{C}(5~\mathrm{K}) =0.975 \pm 0.007$ and $\mathcal{C}(20~\mathrm{K}) =0.594 \pm 0.007$. This is in reasonable agreement with the independently measured single-photon indistinguishability   $M_\mathrm{s} = (92.6 \pm 0.1)\%$ at 5~K, and $M_\mathrm{s} = (58.0 \pm 1.0)\%$ at 20~K. {The unitary energy is still maximum for $\theta=\pi/2$ but its maximum value has reduced from  $E^\mathrm{q,f}_\mathrm{unit}/\hbar \omega_0=( 27.9 \pm 1.2)\%$ at 5~K to $E^\mathrm{q,f}_\mathrm{unit}/\hbar \omega_0=(15.8\pm0.6)\%$ at 20~K. These observations attest for the phonon bath-induced loss of coherence of the quantum emitter}.

In the second step of our protocol, we couple the emitted photonic field with a coherent (classical) field on a 50:50 beam splitter (see Fig.~\ref{fig:1}(b)). This standard configuration to obtain effective light-light interaction in optical quantum technologies leads to the same splitting in the energy transfers between light modes: unitary versus correlation (see Suppl.). A transfer of unitary energy corresponds to a change in the coherent energy of the receiving field.  We set the input energies of both fields to be equal to maximize the interference visibility $v$ i.e. {${\cal E}_\mathrm{f}^\mathrm{in}={\cal E}_\mathrm{c}^\mathrm{in}=\hbar \omega_0 \sin^2(\theta/2)$}, {and consider one of the outputs as hosting the resulting field}. {Contrary to the previous scenario where the interference visibility gave direct access to the unitary energy, our calculations show that, in the case of a classical field interfering with a quantum field,  the interference visibility gives now access to the total energy exchanged (see Suppl.). The energy transferred to the classical field now reads $\Delta {\cal E}_\mathrm{f,c}=\mathcal{E}^\mathrm{in}_\mathrm{f}\times v$} with $v=1$ signaling a complete transfer. The visibility is given by $v=\mathcal{C}_\mathrm{f,c}\cos(\theta/2)$ where $\mathcal{C}_\mathrm{f,c}$ quantifies the overall classical and quantum coherence between both fields. In the case where {the QD is initially subject to pure dephasing}, we show that $\mathcal{C}_\mathrm{f,c} = {M_\text{f,c}}$, where $M_\text{f,c}$ is the mean wave packet overlap between the quantum and the classical field.

We find that the correlation energy is dictated by the quantum field (see Suppl.): $E^\mathrm{f,c}_\mathrm{corr}=E^\mathrm{q,f}_\mathrm{corr}/2$ which is consistent with the fact that the quantum field is the only field carrying an incoherent component. As a result, we can deduce the unitary energy component according to $E^\mathrm{f,c}_\mathrm{unit}=\mathcal{E}^\mathrm{in}_\mathrm{f} \times v- E^\mathrm{q,f}_\mathrm{corr}/2$, which is upper-bounded by the unitary energy initially given to the quantum field: $E^\mathrm{q,f}_\mathrm{unit}\geq E^\mathrm{f,c}_\mathrm{unit}$ (see Suppl.). Fig.~\ref{fig:3}(a) presents the theory curves for the unitary, correlation and total energy transfers in the ideal situation of a pure quantum state. Most of the energy is transferred in the form of unitary energy in the limit where $\theta\rightarrow 0$, i.e. when the state of the quantum field is the closest to a coherent field and the energy transfer is complete. $\Delta \mathcal{E}_{\mathrm{f,c}}$ is maximum for $\pi/2<\theta < \pi$, a behavior that results from a trade-off between the maximization of the quantum field coherence reached at $\theta=\pi/2$ while the coherence of the classical field continuously increases with $\theta$.  For high values of $\theta$, the unitary energy becomes negative, {signaling flow of unitary energy in the opposite direction: from the classical field to the quantum field.} Finally, when $\theta=\pi$ the quantum field consists of a single-photon pulse, giving rise to a maximum correlation energy input of half a photon into the classical field. No interference takes place, hence, no energy is transferred to the classical field. In this situation, correlation and unitary energy transfers perfectly cancel out.

We experimentally study this interference both at  5~K  and 20~K  using the experimental configuration 2 in Fig.~\ref{fig:1}(c). To minimize the effect of vibrations in our closed-cycle cryostation, the classical field is sent into the same cryostation and focused by an objective lens (L) onto a mirror. To further increase the signal-to-noise ratio, the classical field is spectrally shaped using Fabry-P\'erot etalons (FP) to increase its temporal overlap with the quantum field (see  Fig.~\ref{fig:temporalprofile}). We observe phase-dependent oscillations in the counts at the two detectors at the output of the beam splitter from which we extract an interference visibility plotted as a function of pulse area $\theta$ in {Fig.~\ref{fig:3}(b)}. As theoretically predicted, an increasing visibility is observed when reducing $\theta$, both at 5~K and 20~K, evidencing energy transfer towards the classical field of increasing efficiency. Our observations are well reproduced  by $v = \Delta\mathcal{E}_\text{f,c}/{\cal E}_\text{f}^\mathrm{in} = \cos(\theta/2)\mathcal{C}_\text{f,c}$ for  $\mathcal{C}_\text{f,c}(5~\mathrm{K}) = (36.3 \pm 0.4) \%$ and $\mathcal{C}_\text{f,c}(20~\mathrm{K})=(27.2 \pm 0.4)\%$.

Fig.~\ref{fig:3}(c) shows the measured unitary, correlation and total energy transfers deduced from the visibility measured in Fig.~\ref{fig:3}(b) and the measurement of the correlation energy received by the quantum field in Fig.~\ref{fig:2}(b). While the energy transfer during spontaneous emission at 5~K was close to the ideal case, this second step deviates from theoretical maximum with significantly reduced maximum energy exchanged, and positive unitary energy flows for smaller $\theta$ range. The situation is even worse at 20~K, where the unitary energy flows the opposite direction for most of the $\theta$ range.

To better quantitatively analyze these observations, we measured the mean wave packet overlaps between the quantum field and the classical field. We find $M_\text{f,c} = (48.9 \pm 0.3)\%$ at 5~K and $M_\text{f,c}= (32.3 \pm 0.7)\%$ at 20~K. These values are lower than expected considering the single-photon mean wave packet overlap and the good temporal overlap of the envelop functions of the both fields (see Fig.~\ref{fig:temporalprofile}). This  could indicate some spectral diffusion of the qubit resonance over long time scales that is not properly captured by the $M_s$ extracted from interfering two quantum fields separated by $12{.5}$~ns only.  Overall, the expected upper bound of $\mathcal{C}_\mathrm{f,c} = {M_\text{f,c}}$ is not reached here, an observation that could be due to some residual blinking of the QD transition that affect differently our measurements of $\mathcal{C}_\mathrm{f,c}$ and ${M_\text{f,c}}$.

In conclusion, we have proposed and implemented experimental protocols to measure the unitary and correlation energy exchanged between a two-level system and the electromagnetic field as well as between two light fields. The unitary energy bears similarities with the still investigated concept of work in quantum thermodynamics \cite{Binder2018}. Thus, our experiment is an important step to measure work exchanges between systems, and not only the work received by a quantum open system. This capacity holds the promise to monitor directly how a battery provides or stores energy,
and to explore how quantum features impacts these exchanges \cite{RMP_Coherence}. This feature has remained elusive so far in experimental studies. Remarkably, it is achieved here without the need to resort to time-resolved reconstruction of  quantum trajectories \cite{Murch2020}. Our study reveals that the maximum unitary energy transfer efficiency is obtained during spontaneous emission at the onset of the qubit population inversion. It also demonstrates how both classical and quantum coherence impact {the nature of} energetic transfers and can be accessed through homodyne measurements.  The two experimental situations studied here constitute key building blocks for a multitude of quantum technologies from atom-based quantum memories, linear optical gates to Bell state measurements among others. The present work thus carries the seeds of an energetic investigation of realistic processes at the core of optical quantum technologies~\cite{Alexia}. \\

\noindent $^{\ast}$Correspondence and requests for data should be addressed to A.A. (alexia.auffeves@cnrs.fr) and P.S. (pascale.senellart-mardon@c2n.upsaclay.fr). \newline

We acknowledge financial support by the Agence Nationale de la Recherche (QuDICE project), the  H2020-FET OPEN project number 899544 - PHOQUSING, the French RENATECH network, the Paris Ile-de-France Region in the framework of DIM SIRTEQ, the Foundational Questions Institute Fund (Grant number FQXi-IAF19-05), the Templeton World Charity Foundation, Inc (Grant No. TWCF0338).

\bibliography{references.bib}

\begin{thebibliography}{10}
\expandafter\ifx\csname url\endcsname\relax
  \def\url#1{\texttt{#1}}\fi
\expandafter\ifx\csname urlprefix\endcsname\relax\def\urlprefix{URL }\fi
\providecommand{\bibinfo}[2]{#2}
\providecommand{\eprint}[2][]{\url{#2}}

\bibitem{Quantum_battery}
\bibinfo{author}{Quach, J.} \emph{et~al.}
\newblock \bibinfo{title}{Superabsorption in an organic microcavity: Toward a
  quantum battery}.
\newblock \emph{\bibinfo{journal}{Science Advances}}
  \textbf{\bibinfo{volume}{8}} (\bibinfo{year}{2022}).

\bibitem{Quantacell}
\bibinfo{author}{Binder, F.~C.}, \bibinfo{author}{Vinjanampathy, S.},
  \bibinfo{author}{Modi, K.} \& \bibinfo{author}{Goold, J.}
\newblock \bibinfo{title}{{Quantacell: Powerful charging of quantum
  batteries}}.
\newblock \emph{\bibinfo{journal}{New J. Phys.}} \textbf{\bibinfo{volume}{17}}
  (\bibinfo{year}{2015}).

\bibitem{Polini2019}
\bibinfo{author}{Andolina, G.~M.} \emph{et~al.}
\newblock \bibinfo{title}{Extractable work, the role of correlations, and
  asymptotic freedom in quantum batteries}.
\newblock \emph{\bibinfo{journal}{Phys. Rev. Lett.}}
  \textbf{\bibinfo{volume}{122}}, \bibinfo{pages}{047702}
  (\bibinfo{year}{2019}).
\newblock
  \urlprefix\url{https://link.aps.org/doi/10.1103/PhysRevLett.122.047702}.

\bibitem{Stevens2022}
\bibinfo{author}{Stevens, J.} \emph{et~al.}
\newblock \bibinfo{title}{Energetics of a single qubit gate}.
\newblock \emph{\bibinfo{journal}{Phys. Rev. Lett.}}
  \textbf{\bibinfo{volume}{129}}, \bibinfo{pages}{110601}
  (\bibinfo{year}{2022}).
\newblock
  \urlprefix\url{https://link.aps.org/doi/10.1103/PhysRevLett.129.110601}.

\bibitem{GB}
\bibinfo{author}{Gea-Banacloche, J.}
\newblock \bibinfo{title}{Minimum energy requirements for quantum computation}.
\newblock \emph{\bibinfo{journal}{Phys. Rev. Lett.}}
  \textbf{\bibinfo{volume}{89}}, \bibinfo{pages}{217901}
  (\bibinfo{year}{2002}).
\newblock
  \urlprefix\url{https://link.aps.org/doi/10.1103/PhysRevLett.89.217901}.

\bibitem{Ozawa}
\bibinfo{author}{Ozawa, M.}
\newblock \bibinfo{title}{Conservative quantum computing}.
\newblock \emph{\bibinfo{journal}{Phys. Rev. Lett.}}
  \textbf{\bibinfo{volume}{89}}, \bibinfo{pages}{057902}
  (\bibinfo{year}{2002}).
\newblock
  \urlprefix\url{https://link.aps.org/doi/10.1103/PhysRevLett.89.057902}.

\bibitem{Alexia}
\bibinfo{author}{Auff\`eves, A.}
\newblock \bibinfo{title}{Quantum technologies need a quantum energy
  initiative}.
\newblock \emph{\bibinfo{journal}{PRX Quantum}} \textbf{\bibinfo{volume}{3}},
  \bibinfo{pages}{020101} (\bibinfo{year}{2022}).
\newblock \urlprefix\url{https://link.aps.org/doi/10.1103/PRXQuantum.3.020101}.

\bibitem{Binder2018}
\bibinfo{author}{Binder, F.}, \bibinfo{author}{Correa, L.~A.},
  \bibinfo{author}{Gogolin, C.}, \bibinfo{author}{Anders, J.} \&
  \bibinfo{author}{Adesso, G.}
\newblock \emph{\bibinfo{title}{Thermodynamics in the quantum regime}}, vol.
  \bibinfo{volume}{195} of \emph{\bibinfo{series}{Fundamental Theories of
  Physics}} (\bibinfo{publisher}{Springer}, \bibinfo{year}{2018}).

\bibitem{Alipour2016}
\bibinfo{author}{Alipour, S.} \emph{et~al.}
\newblock \bibinfo{title}{{Correlations in quantum thermodynamics: Heat, work,
  and entropy production}}.
\newblock \emph{\bibinfo{journal}{Sci. Rep.}} \textbf{\bibinfo{volume}{6}},
  \bibinfo{pages}{1--14} (\bibinfo{year}{2016}).
\newblock \urlprefix\url{http://dx.doi.org/10.1038/srep35568}.
\newblock \eprint{1606.08869}.

\bibitem{Hossein-Nejad2015}
\bibinfo{author}{Hossein-Nejad, H.}, \bibinfo{author}{O'Reilly, E.~J.} \&
  \bibinfo{author}{Olaya-Castro, A.}
\newblock \bibinfo{title}{Work, heat and entropy production in bipartite
  quantum systems}.
\newblock \emph{\bibinfo{journal}{New J. Phys.}} \textbf{\bibinfo{volume}{17}},
  \bibinfo{pages}{075014} (\bibinfo{year}{2015}).
\newblock
  \urlprefix\url{https://iopscience.iop.org/article/10.1088/1367-2630/17/7/075014}.

\bibitem{Weimer2008}
\bibinfo{author}{Weimer, H.}, \bibinfo{author}{Henrich, M.~J.},
  \bibinfo{author}{Rempp, F.}, \bibinfo{author}{Schr\"{o}der, H.} \&
  \bibinfo{author}{Mahler, G.}
\newblock \bibinfo{title}{Local effective dynamics of quantum systems: A
  generalized approach to work and heat}.
\newblock \emph{\bibinfo{journal}{EPL}} \textbf{\bibinfo{volume}{83}},
  \bibinfo{pages}{30008} (\bibinfo{year}{2008}).
\newblock
  \urlprefix\url{https://iopscience.iop.org/article/10.1209/0295-5075/83/30008}.

\bibitem{Schroder2018}
\bibinfo{author}{Schr\"{o}der, H.} \& \bibinfo{author}{Mahler, G.}
\newblock \bibinfo{title}{Work exchange between quantum systems: The
  spin-oscillator model}.
\newblock \emph{\bibinfo{journal}{Phys. Rev. E}} \textbf{\bibinfo{volume}{81}},
  \bibinfo{pages}{021118} (\bibinfo{year}{2010}).
\newblock \urlprefix\url{https://link.aps.org/doi/10.1103/PhysRevE.81.021118}.

\bibitem{Ciccarello2021}
\bibinfo{author}{Ciccarello, F.}, \bibinfo{author}{Lorenzo, S.},
  \bibinfo{author}{Giovannetti, V.} \& \bibinfo{author}{Palma, G.~M.}
\newblock \bibinfo{title}{{Quantum collision models: open system dynamics from
  repeated interactions}}.
\newblock \emph{\bibinfo{journal}{arXiv}} \textbf{\bibinfo{volume}{2106.11974}}
  (\bibinfo{year}{2021}).

\bibitem{fan2010}
\bibinfo{author}{Fan, S.}, \bibinfo{author}{Kocaba\c{s}, {\c{S}}.~E.} \&
  \bibinfo{author}{Shen, J.-T.}
\newblock \bibinfo{title}{Input-output formalism for few-photon transport in
  one-dimensional nanophotonic waveguides coupled to a qubit}.
\newblock \emph{\bibinfo{journal}{Phys. Rev. A}} \textbf{\bibinfo{volume}{82}},
  \bibinfo{pages}{063821} (\bibinfo{year}{2010}).
\newblock \urlprefix\url{https://link.aps.org/doi/10.1103/PhysRevA.82.063821}.

\bibitem{landi2019}
\bibinfo{author}{Rodrigues, F.}, \bibinfo{author}{De~Chiara, G.},
  \bibinfo{author}{Paternostro, M.} \& \bibinfo{author}{Landi, G.}
\newblock \bibinfo{title}{Thermodynamics of weakly coherent collisional
  models}.
\newblock \emph{\bibinfo{journal}{Physical Review Letters}}
  \textbf{\bibinfo{volume}{123}} (\bibinfo{year}{2019}).

\bibitem{Andolina2018}
\bibinfo{author}{Andolina, G.~M.} \emph{et~al.}
\newblock \bibinfo{title}{Charger-mediated energy transfer in exactly solvable
  models for quantum batteries}.
\newblock \emph{\bibinfo{journal}{Phys. Rev. B}} \textbf{\bibinfo{volume}{98}},
  \bibinfo{pages}{205423} (\bibinfo{year}{2018}).
\newblock \urlprefix\url{https://link.aps.org/doi/10.1103/PhysRevB.98.205423}.

\bibitem{Ferraro2018}
\bibinfo{author}{Ferraro, D.}, \bibinfo{author}{Campisi, M.},
  \bibinfo{author}{Andolina, G.~M.}, \bibinfo{author}{Pellegrini, V.} \&
  \bibinfo{author}{Polini, M.}
\newblock \bibinfo{title}{{High-Power Collective Charging of a Solid-State
  Quantum Battery}}.
\newblock \emph{\bibinfo{journal}{Phys. Rev. Lett.}}
  \textbf{\bibinfo{volume}{120}}, \bibinfo{pages}{117702}
  (\bibinfo{year}{2018}).
\newblock \urlprefix\url{https://doi.org/10.1103/PhysRevLett.120.117702}.
\newblock \eprint{1707.04930}.

\bibitem{Monsel2020}
\bibinfo{author}{Monsel, J.}, \bibinfo{author}{Fellous-Asiani, M.},
  \bibinfo{author}{Huard, B.} \& \bibinfo{author}{Auff{\`{e}}ves, A.}
\newblock \bibinfo{title}{{The Energetic Cost of Work Extraction}}.
\newblock \emph{\bibinfo{journal}{Phys. Rev. Lett.}}
  \textbf{\bibinfo{volume}{124}}, \bibinfo{pages}{1--6} (\bibinfo{year}{2020}).
\newblock \eprint{1907.00812}.

\bibitem{Maffei2021}
\bibinfo{author}{Maffei, M.}, \bibinfo{author}{Camati, P.~A.} \&
  \bibinfo{author}{Auff\`eves, A.}
\newblock \bibinfo{title}{Probing nonclassical light fields with energetic
  witnesses in waveguide quantum electrodynamics}.
\newblock \emph{\bibinfo{journal}{Phys. Rev. Research}}
  \textbf{\bibinfo{volume}{3}}, \bibinfo{pages}{L032073}
  (\bibinfo{year}{2021}).
\newblock
  \urlprefix\url{https://link.aps.org/doi/10.1103/PhysRevResearch.3.L032073}.

\bibitem{Mikko2017}
\bibinfo{author}{Ikonen, J., J.and~Salmilehto} \& \bibinfo{author}{Möttönen,
  M.}
\newblock \bibinfo{title}{Energy-efficient quantum computing}.
\newblock \emph{\bibinfo{journal}{npj Quantum Inf}}
  \textbf{\bibinfo{volume}{3}}, \bibinfo{pages}{17} (\bibinfo{year}{2017}).

\bibitem{cimini2020}
\bibinfo{author}{Cimini, V.} \emph{et~al.}
\newblock \bibinfo{title}{Experimental characterization of the energetics of
  quantum logic gates}.
\newblock \emph{\bibinfo{journal}{npj Quantum Information}}
  \textbf{\bibinfo{volume}{6}}, \bibinfo{pages}{96} (\bibinfo{year}{2020}).
\newblock \urlprefix\url{https://doi.org/10.1038/s41534-020-00325-7}.

\bibitem{VonLindenfels2019}
\bibinfo{author}{{Von Lindenfels}, D.} \emph{et~al.}
\newblock \bibinfo{title}{{Spin Heat Engine Coupled to a Harmonic-Oscillator
  Flywheel}}.
\newblock \emph{\bibinfo{journal}{Phys. Rev. Lett.}}
  \textbf{\bibinfo{volume}{123}}, \bibinfo{pages}{80602}
  (\bibinfo{year}{2019}).
\newblock \urlprefix\url{https://doi.org/10.1103/PhysRevLett.123.080602}.
\newblock \eprint{1808.02390}.

\bibitem{Peterson2019}
\bibinfo{author}{Peterson, J.~P.} \emph{et~al.}
\newblock \bibinfo{title}{{Experimental Characterization of a Spin Quantum Heat
  Engine}}.
\newblock \emph{\bibinfo{journal}{Phys. Rev. Lett.}}
  \textbf{\bibinfo{volume}{123}}, \bibinfo{pages}{240601}
  (\bibinfo{year}{2019}).
\newblock \urlprefix\url{https://doi.org/10.1103/PhysRevLett.123.240601}.
\newblock \eprint{1803.06021}.

\bibitem{Cottet2017}
\bibinfo{author}{Cottet, N.} \emph{et~al.}
\newblock \bibinfo{title}{{Observing a quantum Maxwell demon at work}}.
\newblock \emph{\bibinfo{journal}{Proc. Natl. Acad. Sci. U. S. A.}}
  \textbf{\bibinfo{volume}{114}}, \bibinfo{pages}{7561--7564}
  (\bibinfo{year}{2017}).
\newblock \eprint{1702.05161}.

\bibitem{giesz2016}
\bibinfo{author}{Giesz, V.} \emph{et~al.}
\newblock \bibinfo{title}{Coherent manipulation of a solid-state artificial
  atom with few photons}.
\newblock \emph{\bibinfo{journal}{Nature Communications}}
  \textbf{\bibinfo{volume}{7}} (\bibinfo{year}{2016}).

\bibitem{Somaschi2016}
\bibinfo{author}{Somaschi, N.} \emph{et~al.}
\newblock \bibinfo{title}{{Near-optimal single-photon sources in the solid
  state}}.
\newblock \emph{\bibinfo{journal}{Nature Photonics}}
  \textbf{\bibinfo{volume}{10}}, \bibinfo{pages}{340--345}
  (\bibinfo{year}{2016}).
\newblock \urlprefix\url{https://doi.org/10.1038/nphoton.2016.23}.

\bibitem{Loredo2019}
\bibinfo{author}{Loredo, J.~C.} \emph{et~al.}
\newblock \bibinfo{title}{Generation of non-classical light in a photon-number
  superposition}.
\newblock \emph{\bibinfo{journal}{Nature Photonics}}
  \textbf{\bibinfo{volume}{13}}, \bibinfo{pages}{803--808}
  (\bibinfo{year}{2019}).
\newblock \urlprefix\url{https://doi.org/10.1038/s41566-019-0506-3}.

\bibitem{Grange2017}
\bibinfo{author}{Grange, T.} \emph{et~al.}
\newblock \bibinfo{title}{Reducing phonon-induced decoherence in solid-state
  single-photon sources with cavity quantum electrodynamics}.
\newblock \emph{\bibinfo{journal}{Phys. Rev. Lett.}}
  \textbf{\bibinfo{volume}{118}}, \bibinfo{pages}{253602}
  (\bibinfo{year}{2017}).
\newblock
  \urlprefix\url{https://link.aps.org/doi/10.1103/PhysRevLett.118.253602}.

\bibitem{santori2002}
\bibinfo{author}{Santori, C.}, \bibinfo{author}{Fattal, D.},
  \bibinfo{author}{Vu{\v c}kovi{\'c}, J.}, \bibinfo{author}{Solomon, G.~S.} \&
  \bibinfo{author}{Yamamoto, Y.}
\newblock \bibinfo{title}{Indistinguishable photons from a single-photon
  device}.
\newblock \emph{\bibinfo{journal}{Nature}} \textbf{\bibinfo{volume}{419}},
  \bibinfo{pages}{594--597} (\bibinfo{year}{2002}).
\newblock \urlprefix\url{https://doi.org/10.1038/nature01086}.

\bibitem{RMP_Coherence}
\bibinfo{author}{Streltsov, A.}, \bibinfo{author}{Adesso, G.} \&
  \bibinfo{author}{Plenio, M.~B.}
\newblock \bibinfo{title}{Colloquium: Quantum coherence as a resource}.
\newblock \emph{\bibinfo{journal}{Rev. Mod. Phys.}}
  \textbf{\bibinfo{volume}{89}}, \bibinfo{pages}{041003}
  (\bibinfo{year}{2017}).
\newblock
  \urlprefix\url{https://link.aps.org/doi/10.1103/RevModPhys.89.041003}.

\bibitem{Murch2020}
\bibinfo{author}{Naghiloo, M.} \emph{et~al.}
\newblock \bibinfo{title}{Heat and work along individual trajectories of a
  quantum bit}.
\newblock \emph{\bibinfo{journal}{Phys. Rev. Lett.}}
  \textbf{\bibinfo{volume}{124}}, \bibinfo{pages}{110604}
  (\bibinfo{year}{2020}).
\newblock
  \urlprefix\url{https://link.aps.org/doi/10.1103/PhysRevLett.124.110604}.

\bibitem{ozdemir2002pulse}
\bibinfo{author}{{\"O}zdemir, {\c{S}}.~K.}, \bibinfo{author}{Miranowicz, A.},
  \bibinfo{author}{Koashi, M.} \& \bibinfo{author}{Imoto, N.}
\newblock \bibinfo{title}{Pulse-mode quantum projection synthesis: Effects of
  mode mismatch on optical state truncation and preparation}.
\newblock \emph{\bibinfo{journal}{Physical Review A}}
  \textbf{\bibinfo{volume}{66}}, \bibinfo{pages}{053809}
  (\bibinfo{year}{2002}).

\bibitem{ollivier2021hong}
\bibinfo{author}{Ollivier, H.} \emph{et~al.}
\newblock \bibinfo{title}{Hong-ou-mandel interference with imperfect single
  photon sources}.
\newblock \emph{\bibinfo{journal}{Physical Review Letters}}
  \textbf{\bibinfo{volume}{126}}, \bibinfo{pages}{063602}
  (\bibinfo{year}{2021}).

\bibitem{Wein2021PhotonnumberEG}
\bibinfo{author}{Wein, S.~C.} \emph{et~al.}
\newblock \bibinfo{title}{Photon-number entanglement generated by sequential
  excitation of a two-level atom}.
\newblock \emph{\bibinfo{journal}{arXiv}} \textbf{\bibinfo{volume}{2106.02049}}
  (\bibinfo{year}{2021}).

\end{thebibliography}
\bibliographystyle{naturemag}

\clearpage
\onecolumngrid

	\appendix
	\setcounter{figure}{0} \renewcommand{\thefigure}{S.\arabic{figure}}
	\setcounter{equation}{0} 
	\renewcommand{\theequation}{S.\arabic{equation}}
	\setcounter{table}{0} 
	\renewcommand{\thetable}{S.\arabic{table}}

{\begin{center}
    \large{Supplementary Material}
\end{center}}

\subsection*{I. Description of the photonic field}
In this section we provide a description of the photonic field state in realistic conditions where decoherence  perturbs the spontaneous emission mechanism. The photonic density operator in the pulse-mode formalism \cite{ozdemir2002pulse} of a single propagating mode $\hat{a}_\mathrm{f}(t)$ containing at most one photon reads:
\begin{equation}
\begin{aligned}
\label{generalDensity}
    \hat{{\rho}}_\mathrm{f} &= p_0\ket{0}\bra{0} + p_1\iint dtdt^\prime\xi(t,t^\prime)\hat{a}_{\mathrm{f}}^\dagger(t)\ket{0}\bra{0}\hat{a}_{\mathrm{f}}(t^\prime)+\sqrt{p_0p_1}\int dt\zeta(t)\hat{a}_{\mathrm{f}}^\dagger(t)\ket{0}\bra{0}+\text{h.c.}\\
    &=p_0\hat{{\rho}}_0 + p_1\hat{{\rho}}_1 + \sqrt{p_0p_1}\left(\hat{{\rho}}_{01}+\hat{{\rho}}_{10}\right),
\end{aligned}
\end{equation}
where $\xi(t,t^\prime)=\xi^*(t^\prime,t)$ is a Hermitian function describing the temporal shape and coherence of the single photon and $\zeta(t)$ is the complex amplitude describing the time dynamics of the photon-number coherence. The photon-number probabilities $p_0$ and $p_1$ satisfy $p_0+p_1=1$ and so $\text{Tr}[\hat{{\rho}}_\mathrm{f}]=1$ implies that $\int\xi(t,t)dt=1$. The total purity of this photonic state is
\begin{equation}
\begin{aligned}
    \mathcal{P}&=\text{Tr}\left[\hat{{\rho}}_\mathrm{f}^2\right]
    =p_0^2 + p_1^2 M_\mathrm{s} + 2p_0p_1\mathcal{C},
\end{aligned}
\end{equation}
where
\begin{equation}
    M_\mathrm{s} = \text{Tr}\left[\hat{{\rho}}_1^2\right]=\iint dtdt^\prime\left|\xi(t,t^\prime)\right|^2
\end{equation}
is the single-photon indistinguishability, or purity in the temporal domain, and
\begin{equation}\label{c1}
    \mathcal{C} = \text{Tr}\left[\hat{{\rho}}_{01}\hat{{\rho}}_{10}\right]=\int dt\left|\zeta(t)\right|^2
\end{equation}
is the number purity of the coherence {between the single photon and} the vacuum.

We can compute the expected values for $M_\text{s}$ and $\mathcal{C}$ for a two-level emitter that is ideally prepared in an initial state, but that emits into the waveguide under the influence of pure dephasing and spectral diffusion. The temporal wavefunction of a single photon produced by a two-level emitter in the absence of pure dephasing is given by $f(t,\omega_0)= \sqrt{\gamma}e^{-\gamma t/2 - i\omega_0t}$, where $\gamma$ is the spontaneous emission rate of the two-level emitter and $\omega_0$ is the transition frequency. Pure dephasing, by definition, causes an exponentially decaying temporal coherence between the photonic state at two instantaneous times $t$ and $t^\prime$. This is captured by the temporal density function and coherence function of the forms
\begin{equation}
\label{temporaldensity}
    \xi(t,t^\prime,\omega_0)= f(t,\omega_0)f^*(t^\prime,\omega_0)e^{-\gamma^\star|t-t^\prime|}\hspace{10mm}\zeta(t,\omega_0) = f(t,\omega_0)e^{-\gamma^\star t}
\end{equation}
One can see that the absolute value of the difference in times $t$ and $t^\prime$ prevents this temporal density function $\xi$ from being factored using some new wavefunction $f^\prime$, hence the single-photon component of the photonic state is no longer temporally pure.

When measuring $M_\text{s}$ or $\mathcal{C}$, two photonic states arising from successive excitations of the emitter are interfered in a path-unbalanced Mach-Zehnder interferometer (see section IV.2). The time delay between excitations allows the emitter to shift emission frequency by some small amount $\delta\omega$, although we still consider the frequency of the transition for each individual emission event to be fixed on the timescale of $1/\gamma$. This spectral diffusion of the emission implies a reduced mean wave packet overlap, and hence a reduced single-photon indistinguishability of
\begin{equation}
    M_\text{s}(\delta\omega) = \iint dtdt^\prime\text{Re}\left[\xi(t,t^\prime,\omega_0)\xi^*(t,t^\prime,\omega_0+\delta\omega)\right].
\end{equation}
Evaluating this expression for the photonic state of the form given by Eq. (\ref{generalDensity}) using the function $\xi$ from Eq. (\ref{temporaldensity}) gives
\begin{equation}
    M_\text{s}(\delta\omega) = \frac{\gamma(\gamma+2\gamma^\star)}{(\gamma+2\gamma^\star)^2+\delta\omega^2}.
\end{equation}
In this expression, we can identify $\gamma+2\gamma^\star=2/T_2$ which is related to the total decoherence rate and gives the homogeneously-broadened line width of the emitter for a single emission. The line is further broadened by a classical averaging over $\delta\omega$ following a distribution describing the spectral diffusion of the emitter.

Similarly, we can find that the number purity of the coherence between the single photon and the vacuum measured in the presence of spectral diffusion is susceptible to the decoherence caused by pure dephasing, but also by the random phase fluctuations caused by the spectral shift of the emitter:
\begin{equation}
\mathcal{C}(\delta\omega) =\int dt \text{Re}\left[\zeta(t,\omega_0)\zeta^*(t,\omega_0+\delta\omega)\right].
\end{equation}
Using our temporal wavefunction solutions for an ideally-prepared two-level emitter, we obtain exactly the same result as for the mean wave packet overlap
\begin{equation}
    \mathcal{C}(\delta\omega) = \frac{\gamma(\gamma+2\gamma^\star)}{(\gamma+2\gamma^\star)^2+\delta\omega^2} = M_\text{s}(\delta\omega).
\end{equation}
Thus, regardless of the distribution of spectral fluctuations, both pure dephasing and spectral diffusion degrade $M_\text{s}$ and $\mathcal{C}$ equivalently.

Note that the result $\mathcal{C}=M_\text{s}$ is not necessarily true for all dephasing effects. In general, the fact that the photonic density matrix must remain positive semi-definite implies that $\mathcal{C} \leq \sqrt{M_\text{s}}$, which arises from the Cauchy-Schwarz inequality and implies that $\mathcal{C}$ may be larger than $M_\text{s}$. This upper bound scaling was observed when varying the polarization of interfering wave packets in Ref. \cite{Loredo2019}, and it held because modifying the polarization in a controlled way does not induce additional phase fluctuations between the vacuum and single-photon states that impact $\mathcal{C}$. Hence, in that work, $\mathcal{C}$ was forced to follow the trend set by the physical upper bound, which is set by the single-photon indistinguishability.

\subsubsection*{II. Effect of the pure dephasing on the energetic analysis}
 
The analysis of the energetics during spontaneous emission builds on the general two-body quantum energetic analysis of~\cite{Alipour2016,Hossein-Nejad2015,Weimer2008} applied to the specific case where the closed two-body system comprises a qubit and a traveling electromagnetic field~\cite{Maffei2021}. Here we show this framework remains valid even when the qubit, while interacting with the electromagnetic field, is coupled to a thermal bath of phonons causing its pure dephasing. In order to show this result, let us first summarize the main results of Refs.~\cite{Alipour2016,Hossein-Nejad2015,Weimer2008} that represent the starting point of our energetic analysis. For any pair of systems (A and B) coupled by an Hamiltonian interaction $V_{A,B}$, the joint state at time t can be written as $\rho(t)=\rho^{A}(t)\otimes\rho^{B}(t)+\chi_{A,B}(t)$, where $\rho^{A(B)}(t) = \text{Tr}_{B(A)}\lbrace \rho(t)\rbrace$ and $\chi_{A,B}(t)$ is the correlation matrix.

The definitions of unitary energy and correlation energy follow naturally from this separation of the dynamics:
\begin{equation}
\dot{{E}}^{A(B)}_\mathrm{unit} (t)=-\frac{i}{\hbar}\text{Tr}_{A,B}\left\lbrace \left[H_{A(B)},V_{A,B}\right]\rho^{A}\left(t\right)\otimes\rho^{B}\left(t\right)\right\rbrace, \label{eq_BQT-W}
\end{equation}
\begin{equation}
    \dot{{E}}^{A(B)}_\mathrm{corr} (t)= -\frac{i}{\hbar}\text{Tr}_{A,B}\left\lbrace \left[H_{A(B)},V_{A,B}\right]\chi_{A,B}(t)\right\rbrace \label{eq_BQT-Q},
\end{equation}
with $H_{A(B)}$ being the bare Hamiltonian of system A(B). For each system the change of the internal energy is given by the sum of unitary and correlation energy, i.e. $\Delta \mathcal{U}_{A(B)}={{E}}^{A(B)}_\mathrm{unit}+{{E}}^{A(B)}_\mathrm{corr}$. 
The two fluxes in unitary energy $\dot{{E}}^{A}_\mathrm{unit}$ and $\dot{{E}}^{B}_\mathrm{unit}$ are connected by the balance relation: \begin{align}\label{eq_balance}\dot{{E}}^{A}_\mathrm{unit}+\dot{{E}}^{B}_\mathrm{unit}-\frac{i}{\hbar}\text{Tr}\lbrace [V_{A,B},H_A +H_B]\rho^{A}(t)\otimes \rho^{B}(t)\rbrace=0.
\end{align}
When A is the qubit and B is its spontaneously emitted field (battery), $\text{Tr}\lbrace [V_{A,B},H_{A}+H_{B}]\rho^{A}(t)\otimes \rho^{B}(t)\rbrace=0$ at any time. This can be shown by replacing the operators in the commutator with their explicit expressions: $H_{A}=\hbar\omega_0 \sigma^{\dagger}\sigma$, $H_{B}=\hbar\sum_{k}\omega_k b^{\dagger}_{k} b_{k}$, $V_{A,B}=\hbar g_0 \sum_{k} (b_k \sigma^{\dagger}+b^{\dagger}_{k} \sigma)$, where $\sigma=\ket{g}\bra{e}$. Injecting these expressions in the commutator we get: $\text{Tr}\lbrace[V_{A,B},H_{A}+H_{B}]\rho^{A}(t)\otimes \rho^{B}(t)\rbrace=\hbar g_0\sum_{k} (\omega_k -\omega_0)\text{Re}\lbrace\langle b_{k}\rangle \langle \sigma\rangle^{*}\rbrace$. When the field is the qubit's spontaneous emission, $\text{Re}\lbrace\langle b_{k}\rangle \langle \sigma\rangle^{*}\rbrace\propto[\gamma^2/4 +(\omega_k-\omega_0)^2]^{-1}$, the spectrum is symmetric around $\omega_0$, and hence summing over $k$, positive and negative contributions cancel each other out. This hypothesis is fulfilled here with the cavity  suppressing phonon sideband emission by several orders of magnitude~\cite{Grange2017}. This implies $\dot{{E}}^{B}_\mathrm{unit}=-\dot{{E}}^{A}_\mathrm{unit}=\hbar \omega_0 |\langle a_{f}(t)\rangle|^2$, where the last equality  has been derived in Ref.~\cite{Maffei2021}.

In order to include pure dephasing in the analysis we have to write down the full Hamiltonian of the three-body system comprising qubit, electromagnetic field, phonons. As there is no direct coupling between the electromagnetic field and the phonons the Hamiltonian will have the form: $H_{A,B,C}=H_{A}+H_{B}+H_{C}+V_{A,B}+V_{A,C}$, where now A=qubit, B=electromagnetic field, and C=phonons. The joint system state can be written as $\rho(t)=\rho^{A}(t)\otimes\rho^{B}(t)\otimes\rho^{C}(t)+\chi_{A,B,C}(t)$. Now the unitary and correlation energy received by A read:
\begin{equation}
\dot{{E}}^{A}_\mathrm{unit}(t)=-\frac{i}{\hbar}\text{Tr}_{A,B}\left\lbrace \left[H_{A},V_{A,B}\right]\rho^{A}\left(t\right)\otimes\rho^{B}\left(t\right)\right\rbrace-\frac{i}{\hbar}\text{Tr}_{A,C}\left\lbrace \left[H_{A},V_{A,C}\right]\rho^{A}\left(t\right)\otimes\rho^{C}\left(t\right)\right\rbrace, \label{eq_BQT-W_A}
\end{equation}
\begin{equation}
    \dot{{E}}^{A}_\mathrm{corr}(t)= -\frac{i}{\hbar}\text{Tr}_{A,B}\left\lbrace \left[H_{A},V_{A,B}\right]\chi_{A,B}(t)\right\rbrace-\frac{i}{\hbar}\text{Tr}_{A,C}\left\lbrace \left[H_{A},V_{A,C}\right]\chi_{A,C}(t)\right\rbrace \label{eq_BQT-Q_A},
\end{equation}
while for the systems B and C we find:
\begin{equation}
\dot{{E}}^{B(C)}_\mathrm{unit}(t)=-\frac{i}{\hbar}\text{Tr}_{A,B(C)}\left\lbrace \left[H_{B(C)},V_{A,B(C)}\right]\rho^{A}\left(t\right)\otimes\rho^{B(C)}\left(t\right)\right\rbrace, \label{eq_BQT-W_BC}
\end{equation}
\begin{equation}
    \dot{{E}}^{B(C)}_\mathrm{corr}(t)= -\frac{i}{\hbar}\text{Tr}_{A,B(C)}\left\lbrace \left[H_{B(C)},V_{A,B(C)}\right]\chi_{A,B(C)}(t)\right\rbrace \label{eq_BQT-Q_BC},
\end{equation}
with $\chi_{A,C}(t)=\text{Tr}_{B}\lbrace \chi(t)\rbrace$ and $\chi_{A,B}(t)=\text{Tr}_{C}\lbrace \chi(t)\rbrace$. For each system the change of the internal energy is still given by the sum of unitary and correlation energy, i.e. $\Delta \mathcal{U}_{A(B)}={{E}}^{A(B)}_\mathrm{unit}+{{E}}^{A(B)}_\mathrm{corr}$. 

Let us now write explicitly the Hamiltonian $V_{A,C}$ coupling qubit and phonons' bath; as we are considering pure dephasing, this interaction reads $V_{A,C}=\sum_{k} (\xi_{k} c_{k}+\xi^{*}_{k} c^{\dagger}_{k})\otimes \sigma_z$. This Hamiltonian commutes with the qubit's bare Hamiltonian, hence, in both Eqs.~\eqref{eq_BQT-W_A} and \eqref{eq_BQT-Q_A}, the last terms vanish and the unitary and correlation energy on the qubit take the same expression as in the zero-temperature case, i.e. $\dot{{E}}^{A}_\mathrm{unit}(t)=-\hbar \omega_0 |\langle a_{f}(t)\rangle|^2$. 
The relation linking the unitary energy on the photonic field with the unitary energy on the qubit is still given by Eq.~\eqref{eq_balance}. The term $\text{Tr}\lbrace [V_{A,B},H_{A}+H_{B}]\rho^{A}(t)\otimes \rho^{B}(t)\rbrace$ is still identical to zero at any time as the spectrum of the emission is still symmetric around $\omega_0$. In particular we are assuming that the qubit's pure dephasing only implies a spectral broadening of the spontaneous emission, leaving the center in $\omega_0$, mathematically this means: $\langle a_{f}(t)\rangle \propto e^{-(\gamma/2 +\gamma^*)t -i\omega_0 t}$. Hence we obtain again $\dot{{E}}^{B}_\mathrm{unit}(t)=-\dot{{E}}^{A}_\mathrm{unit}(t)=\hbar \omega_0 |\langle a_{f}(t)\rangle|^2$ as in the zero-temperature case.

\subsection*{III. Relating energetic and optical quantities}

In this section, we {propose an experimental protocol to measure the energy exchanges between the qubit and the electromagnetic field as well as between the photonic field and the classical field}. We demonstrate that homodyne measurements, where two optical fields interfere on a balanced beam splitter (see Fig.~\ref{fig:1}), give direct access to the unitary energy exchanged between two fields. 
For two input fields impinging a balanced beam splitter, the transformation is given by: 

\begin{equation}\label{eq:BSrel}
    \begin{pmatrix} \hat{a}_3(t) \\  \hat{a}_4(t)  \end{pmatrix} = \frac{1}{\sqrt{2}} \begin{pmatrix}
    1 &e^{i\phi} \\-e^{i\phi} & 1 \end{pmatrix} \begin{pmatrix} \hat{a}_1(t) \\  \hat{a}_2(t)  \end{pmatrix} 
\end{equation}
The visibility of interference is defined as:
\begin{equation}
v = \frac{\mu_3 - \mu_4}{\mu_3 + \mu_4}, \label{eq:VSHD}
\end{equation}
\noindent where $\mu_j = \int \langle \hat{a}^\dagger_j(t) \hat{a}_j(t) \rangle dt $ is the average photon number detected in mode $j$ at the output of the beam splitter. 

In what follows, we successively study the case of a self-homodyne experiment, where the input fields are copies of the photonic field of Eq.~\eqref{generalDensity}, i.e. $a_1(t)=a_{\mathrm{f1}}(t)$, and  $a_2(t)=a_{\mathrm{f2}}(t)$, and the case of a standard homodyne experiment where the input fields are the photonic field (f) and the classical field (c), i.e. $a_1(t) = a_\text{c}^\mathrm{in}(t)$, $a_2(t) = a_\mathrm{f}^\mathrm{in}(t)$, $a_3(t) = a_\text{c}^\mathrm{out}(t)$, $a_4(t) = a_\mathrm{f}^\mathrm{out}(t)$.

\subsubsection*{III.1. Measuring the efficiency of energetic transfers in spontaneous emission with self-homodyne interference}

In Refs.~\cite{Monsel2020,Maffei2021}, it has been demonstrated that when the qubit-field system is isolated, the total unitary energy provided by the qubit to the field's mode can be extracted from the final field's state:
\begin{align}\label{generalWork}
    E_\mathrm{unit}^\mathrm{q,f}=\hbar \omega_0\int dt|\langle \hat{a}_{\mathrm{f}}(t)\rangle|^2
\end{align}

We extended the definition above to the experimental situation where the joint qubit-field state may not be perfectly pure. By evaluating Eq.~\eqref{generalWork} on the general state in Eq.~\eqref{generalDensity} with $p_0=\cos^2{\left(\theta/2\right)}$, and $p_1=\sin^2{\left(\theta/2\right)}$, we find $E_\mathrm{unit}^\mathrm{q,f}=\hbar \omega_0 \cos^2{\left(\theta/2\right)} \sin^2{\left(\theta/2\right)} \int dt |\zeta(t)|^2$.

\noindent Now using Eq.~\eqref{c1}, we find the expression for the unitary energy presented in the main text, $E_\mathrm{unit}^\mathrm{q,f}=\hbar \omega_0 \cos^2{\left(\theta/2\right)} \sin^2{\left(\theta/2\right)} \mathcal{C}$.

$\mathcal{C}$ {can be experimentally accessed from the visibility} of the self-homodyne interference {where} two copies of the emitted photonic field are incident on a balanced beam splitter. The two input states are identical except for a relative phase $\phi$ on the number coherence between them which is due to the difference in path length. 

\noindent The numerator of Eq.~\eqref{eq:VSHD} is given by:
\begin{equation}
    \begin{aligned}
    2\mathrm{Re} \left [\int dt  \langle e^{i\phi} \hat{a}^{\dagger}_\mathrm{f1}(t)\hat{a}_\mathrm{f2}(t) \rangle \right ] =2\mathrm{Re} \left [\int dt  e^{i\phi}\langle  \hat{a}^{\dagger}_\mathrm{f1}(t)\rangle \langle \hat{a}_\mathrm{f2}(t) \rangle \right ]= 2 \mathrm{cos} (\phi)\int dt | \langle \hat{a}_\mathrm{f}(t) \rangle | ^2 
\end{aligned}
\end{equation}
 where the first equality comes from the fact that the states of modes $\mathrm{f1}$ and $\mathrm{f2}$ are uncorrelated, the second from the fact that the states are identical, i.e. $\langle\hat{a}_\mathrm{f1}\rangle=\langle\hat{a}_\mathrm{f2}\rangle=\langle\hat{a}_\mathrm{f}\rangle $.
 
 \noindent Since the two fields carry the same number of photons, the denominator of Eq.~\eqref{eq:VSHD} is simply $2 \mu_\mathrm{f}=2\mathcal{E}_{\mathrm{f}}/(\hbar \omega_0)$. Putting all together we get:

\begin{equation}
    v= \mathrm{cos} (\phi) \frac{\int dt | \langle \hat{a}_{\mathrm{f}}(t) \rangle | ^2}{\mu_\mathrm{f}}=\mathrm{cos} (\phi) \frac{E_\mathrm{unit}^\mathrm{q,f}}{\mathcal{E}_\mathrm{f}}
\end{equation}
 
The visibility of the interference varies with the relative phase between the two inputs, $\phi$, and reaches maximum constructive (destructive) interference when $\phi = 0 \, (\pi)$.\\

\subsubsection*{III.2.  Transfer of energy from the photonic field to the classical field through homodyne interference}

\noindent We transfer energy from the photonic field via interference to a coherent field. The visibility of this classical-homodyne measurement allows us to quantify the energy and the unitary energy transfer. The photonic field (f) enters the balanced beam splitter in channel 2, with $\langle \hat{a}^\mathrm{in}_{\mathrm{f}}(t) \rangle  = \sin{\left(\theta/2\right)}\cos{\left(\theta/2\right)} \zeta (t)$. The classical field (c) field, $\ket{\beta}$, enters the beam splitter in channel 1, with $\langle \hat{a}^\mathrm{in}_{\text{c}}(t) \rangle= \beta(t)$.

\noindent {The final energy of the classical receiver field, reads:}
\begin{align}
  \mathcal{E}^\mathrm{out}_\text{c}&=\frac{1}{2}\left[\mathcal{E}^\mathrm{in}_\text{c}+\mathcal{E}^\mathrm{in}_\mathrm{f}\right]+\hbar\omega_0\mathrm{Re} \left [ \int dt \langle \hat{a}^\mathrm{in}_\mathrm{f}(t) \rangle \langle \hat{a}^\mathrm{in}_\text{c}(t) \rangle ^*\right]\\ \nonumber
  &=\frac{\hbar\omega_0}{2}\left[\int dt |\beta(t)|^2+\sin^2{(\theta/2)}\right]+\hbar\omega_0\cos{(\theta/2)}\sin{(\theta/2)}\mathrm{Re}\left [ \int dt \zeta(t) \beta(t)^*\right]
\end{align}
The {efficiency} of the energy transfer reads $G=\mathcal{E}^\mathrm{out}_\text{c}/\left[\hbar\omega_0\int dt |\beta(t)|^2+\hbar\omega_0\sin^{2}{\left(\theta/2\right)}\right]$, replacing the numerator with the expression above, we find that $G$ is maximal when $\int dt |\beta(t)|^2=\sin^{2}{\left(\theta/2\right)}$, i.e. $\beta(t)=\sin{(\theta/2)}\sqrt{\xi(t,t)}e^{-i\phi(t)}$, namely the two fields are matched in intensity. 
This condition is met in the experiment presented in the main and it corresponds to an interference visibility reading:
\begin{equation}\label{eq:VHD}
    v = \frac{1}{\mu_\mathrm{f}} \mathrm{Re} \left [ \int dt \langle \hat{a}^\mathrm{in}_{\text{c}}(t) \rangle \langle \hat{a}^\mathrm{in}_\mathrm{f}(t) \rangle ^*\right] = \cos{\left(\theta/2\right)} \mathrm{Re} \left[ \int dt \zeta(t) \sqrt{\xi(t,t)}e^{i\phi(t)}\right]=\cos{\left(\theta/2\right)}\mathcal{C}_\text{f,c}
\end{equation}
where, in the last equality, we defined the quantity $\mathcal{C}_\text{f,c}=\text{Re}\left[\int dt \zeta(t) \sqrt{\xi(t,t)}e^{i\phi(t)}\right]$  accounting for both the classical and the quantum coherence of the process. It is clear that the visibility and hence the {relative efficiency} $G=(1+v)/2$ are maximal when $\mathcal{C}_\text{f,c}=1$.
The energy change of the two fields hence reads:
\begin{align}
    \Delta \mathcal{E}_\text{f,c}=\hbar\omega_0\cos{(\theta/2)}\sin^2{(\theta/2)}\mathcal{C}_{\text{f,c}}=-\Delta \mathcal{E}_\mathrm{f}
\end{align}

\noindent We can split this energy change into unitary and correlation energy terms:
\begin{align}
    &\Delta \mathcal{E}_{\mathrm{f,c}}=E^{\text{f,c}}_\mathrm{unit}+E^{\text{f,c}}_\mathrm{corr}
\end{align}

\noindent Using the analysis introduced in Refs.~\cite{Alipour2016,Hossein-Nejad2015,Weimer2008,Schroder2018,Maffei2021}, we find the unitary energy:

\begin{align}\label{eq:work_CW}
    E^{\text{f,c}}_\mathrm{unit}&=\hbar\omega_0\left(\int dt |\langle \hat{a}^\mathrm{out}_\text{c}(t)\rangle|^2-\int dt |\langle \hat{a}^\mathrm{in}_\text{c}(t)\rangle|^2\right)\\ \nonumber
    &=\hbar \omega_0 \mathrm{Re} \left [ \int dt \langle \hat{a}^\mathrm{in}_{\text{c}}(t) \rangle \langle \hat{a}^\mathrm{in}_\mathrm{f}(t) \rangle ^*\right] +\frac{\hbar\omega_0}{2}\left(\int dt |\langle \hat{a}^\mathrm{in}_\mathrm{f}(t)\rangle|^2-\int dt |\langle \hat{a}^\mathrm{in}_\text{c}(t)\rangle|^2\right)\\ \nonumber
    &=\hbar \omega_0\sin^2{(\theta/2)}\left[\cos{(\theta/2)}\mathcal{C}_{\text{f,c}}+\left(\cos^2{(\theta/2)}\mathcal{C} -1\right)/2 \right]
\end{align}
It can be easily verified that for any choice of $\langle \hat{a}^\mathrm{in}_\text{c}(t)\rangle$ we find $ E^{\text{f,c}}_\mathrm{unit}\leq  E^{\text{q,f}}_\mathrm{unit}$. 

\noindent Subtracting the unitary energy from the total energy change we find the correlation energy:
\begin{align}
    E^{\text{f,c}}_\mathrm{corr}=\hbar\omega_0\sin^2{(\theta/2)}\left(1-\cos^2{(\theta/2)}\mathcal{C}\right)/2=\left[\mathcal{E}^\mathrm{in}_{\text{f}}-\hbar\omega_0 \int dt |\langle \hat{a}^\mathrm{in}_\mathrm{f}(t)\rangle|^2\right]/2=E^{\text{q,f}}_\mathrm{corr}/2 
    \end{align}
    
{Similar to the self-homodyne interference, we can derive the explicit results for $\mathcal{C}_\mathrm{f,c}$ and $M_\mathrm{f,c}$ under the effects of pure dephasing and spectral diffusion, assuming a perfectly-prepared two-level emitter. In this case, we must interfere the photonic state from the quantum dot described by $\xi$ and $\zeta$ with the classical field described by the amplitude $\beta(t)=\sin(\theta/2)\sqrt{\xi(t,t)}e^{-i\phi(t)}$. Thus,
\begin{equation}
    M_\mathrm{f,c}(\delta\omega) = \iint dt dt^\prime \text{Re}\left[\sqrt{\xi(t,t,\omega_0)\xi(t^\prime,t^\prime,\omega_0)}e^{-i\omega_0(t-t^\prime)}\xi^*(t,t^\prime,\omega_0+\delta\omega)\right] = \frac{\gamma(\gamma+\gamma^\star)}{(\gamma+\gamma^\star)^2+\delta\omega^2}.
\end{equation}
In this case, the factor $\sin(\theta/2)$ from $\beta(t)$ is canceled in the normalization and we take $\phi(t)=\omega_0 t$. First, we can notice the similarity between this expression and $M_\text{s}(\delta\omega)$. The only difference is that the effect of pure dephasing via $\gamma^\star$ is halved compared to the self-homodyne scenario due to the fact that the classical field does not experience pure dephasing. Second, we can also solve for
\begin{equation}
    \mathcal{C}_\mathrm{f,c}(\delta\omega) = \int dt \text{Re}\left[\sqrt{\xi(t,t,\omega_0)}e^{-i\omega_0 t}\zeta^*(t,t,\omega_0+\delta\omega)\right] = \frac{\gamma(\gamma+\gamma^\star)}{(\gamma+\gamma^\star)^2+\delta\omega^2}=M_\mathrm{f,c}(\delta\omega).
\end{equation}
Thus, just like in the self-homodyne case, we find that the values $M_\mathrm{f,c}$ and $\mathcal{C}_\mathrm{f,c}$ extracted from homodyne interference are equally degraded by pure dephasing and spectral diffusion. Also similar to the self-homodyne case, we can again note that $\mathcal{C}_\mathrm{f,c}=M_\mathrm{f,c}$ is not a general result for every model since a Cauchy-Schwarz inequality will imply a general bound of $\mathcal{C}_\mathrm{f,c}\leq \sqrt{M_\mathrm{f,c}}$. This more general upper bound may be observed, for example, by rotating the polarization of one of the two inputs.}

\subsection*{IV. Experimental methods}

\subsubsection*{IV.1. Photonic field emission}

In the main text, we assumed near unity population inversion both at 5~K and at 20~K at $\pi$ pulse for the sake of simplicity in our data analysis. We show here that this hypothesis is verified a posteriori. Indeed, the self-homodyne measurement also allows us to measure the excited state population probability $p_1$ from the visibility of interference. We can deduce the excited state probability at $\theta=\pi$ according to
\begin{equation}
    p_1(\pi) = 1-\frac{v(\pi)}{\mathcal{C}}
    \label{eq:p1_rabi}
\end{equation}
where we extract $\mathcal{C}$ from the fit in Fig.~\ref{fig:2}(a): $v(\theta)=\mathcal{C}\cos^2{(\theta/2)}$. From this we deduce the values $p_1(\pi)=0.95 \pm 0.19$ at 5K and $p_1(\pi)=0.92 \pm 0.02$ at 20K.

\begin{figure}
	\centering
	\includegraphics[width=0.9\linewidth]{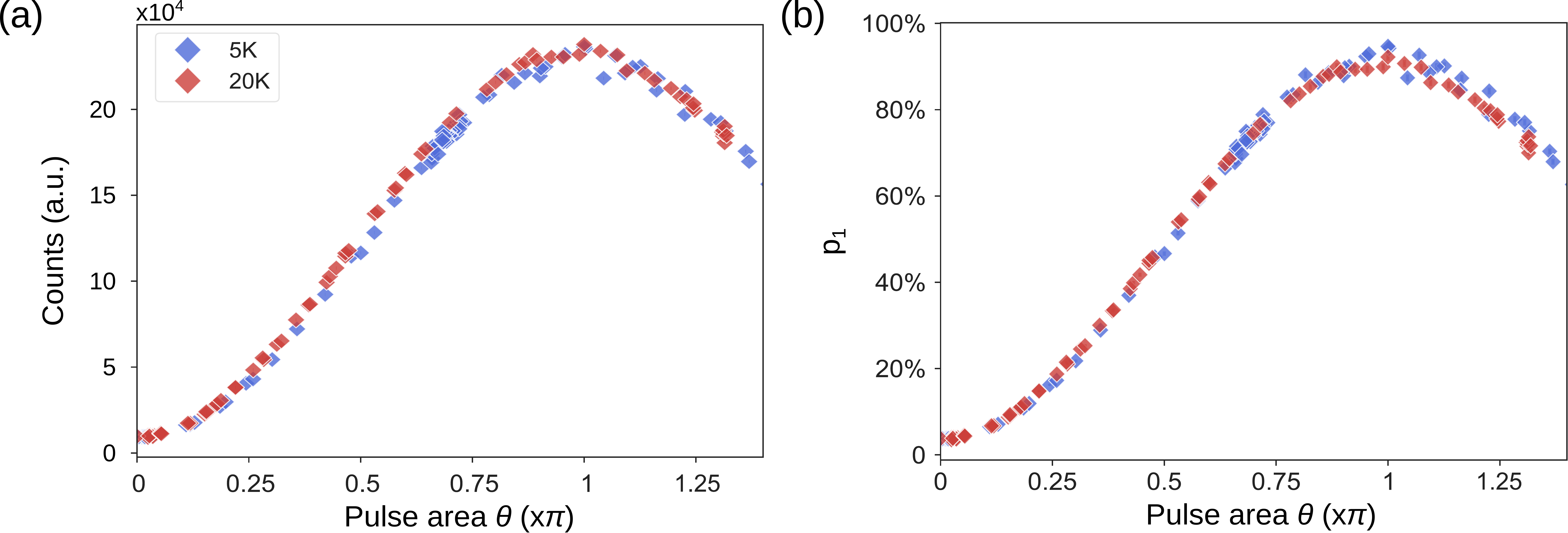}
	\caption{{\textbf{Population excited state} (a) The qubit emission intensity as a function of pulse area $\theta$ at 5~K (blue) and 20~K (red). (b) The corresponding excited state population as a function of pulse area for the qubit performing at 5~K (blue) and at 20~K (red).}  \label{fig:rabipop}}
	\end{figure}
 In Fig.~\ref{fig:rabipop}(a) we show the qubit raw emission intensity measured both at 5~K and 20~K which shows very similar trend - evidencing a very limited effect on the coherence imprinted on the qubit. Fig.~\ref{fig:rabipop}(b) shows the corresponding occupation of the qubit excited state $p_1(\theta)$ using the above measured values $p_1(\pi)$.

\subsubsection*{IV.2. Photonic field indistinguishability}

To measure $\mathcal{C}$ and $M_\mathrm{s}$ for the emitted photonic field, two sequentially generated copies of the state interfere at a 50:50 beam splitter. {We monitor the single-photon intensity at the two outputs of the beam splitter to observe interference fringes, and also simultaneously monitor the two-photon coincidences where we observe bunching due to Hong-Ou-Mandel interference}~\cite{Loredo2019,ollivier2021hong,Wein2021PhotonnumberEG}.
To obtain the single-photon indistinguishability $M_\mathrm{s}$, we first measure the mean wave packet overlap between two subsequently emitted photonic fields, $M$ \cite{ollivier2021hong}. The {mean wave packet overlap} is defined by $M = (1/\mu_\mathrm{f}^2)\iint dtdt^\prime\left|G^{(1)}(t,t^\prime)\right|^2$, where $\mu_\mathrm{f}=\int dtI_\mathrm{f}(t)=\sum_n np_n$ is the average photon number, $I_\mathrm{f}(t)=\braket{\hat{a}_\mathrm{f}^\dagger(t)\hat{a}_\mathrm{f}(t)}$ is the wave packet temporal envelop, and $G^{(1)}(t,t^\prime)=\braket{\hat{a}_\mathrm{f}^\dagger(t^\prime)\hat{a}_\mathrm{f}(t)}$ is the first-order (amplitude) correlation function.
Experimentally, the integrated coincident counts after HOM interference binned with respect to the detection delay $\tau$ produce histograms proportional to $\int dt G^{(2)}_\text{HOM}(t,t+\tau)$, where $2G^{(2)}_\text{HOM}(t,t^\prime)=I_\mathrm{f}(t)I_\mathrm{f}(t^\prime)+G^{(2)}(t,t^\prime)-\left|G^{(1)}(t,t^\prime)\right|^2$ is the phase-averaged intensity correlation after HOM interference and $G^{(2)}(t,t^\prime)=\braket{\hat{a}_\mathrm{f}^\dagger(t)\hat{a}_\mathrm{f}^\dagger(t^\prime)\hat{a}_\mathrm{f}(t^\prime)\hat{a}_\mathrm{f}(t)}$ is the intensity autocorrelation measured using a Hanbury Brown--Twiss setup.

{The required normalization $\mu_\mathrm{f}^2$ for $M$ is obtained by comparing the coincident counts for when the inputs are co-polarized to when they are cross-polarized. In this latter case, we have $G^{(1)}=0$. Then, if the fully integrated and normalized intensity correlation $g^{(2)}=(1/\mu_\mathrm{b}^2)\iint dtdt^\prime G^{(2)}(t,t^\prime)$ is very small, the HOM histograms allow us to quantify $M$ via the HOM visibility defined as:
\begin{equation}
\label{eqvhom}
    {V_\text{HOM} =} \frac{g^{(2)}_{\mathrm{HOM},\perp}-g^{(2)}_{\mathrm{HOM},\parallel}}{g^{(2)}_{\mathrm{HOM},\perp}},
\end{equation}
where $g^{(2)}_{\mathrm{HOM},\parallel} =A_{\tau_0, \parallel}/\overline{A}_{\tau_{>1}, \parallel}$ and $g^{(2)}_{\mathrm{HOM},\perp} =A_{\tau_0, \perp}/\overline{A}_{\tau_{>0}, \perp}$ are the central peak areas ($A_{\tau_0}$) in co- ($\parallel$) and cross- ($\perp$) input polarization configurations normalized by the average peak area of the histogram peaks arising from uncorrelated counts ($\overline{A}_{\tau>1,\parallel}$ or $\overline{A}_{\tau>0,\perp}$). Note that, due to the interferometer delay needed to interfere subsequently generated states, the first side peak of the $\parallel$ case is partially suppressed due to anti-bunching and hence is excluded from the normalization. In addition, Eq.~(\ref{eqvhom}) is an accurate measurement of $M$ only when there is not much first-order coherence in the number basis so that $\overline{A}_{\tau_{>0}, \parallel}$ is not suppressed by possible interference fringes. Hence, this measurement approach is accurate for pulse areas $\theta\simeq\pi$; however, $M$ is expected to take similar values for all $\theta$.} 

{If the mode contains no more than one photon, then $G^{(1)}(t,t^\prime)=\mu_\mathrm{f}^2\xi(t,t^\prime)$ and so $V_\mathrm{HOM}=M=M_\mathrm{s}$. For small nonzero $g^{(2)}$, both $A_{\tau_0,\parallel}$ and $A_{\tau_0,\perp}$ are increased equally, which increases the denominator in Eq. (\ref{eqvhom}) and causes $V_\text{HOM}$ to underestimate $M$. This small underestimate of $M$ can be corrected by taking $M = V_\text{HOM}(1+g^{(2)})$. Note that this measurement approach and subsequent $g^{(2)}$ correction are different than those used in Ref. \cite{ollivier2021hong}, where the visibility was defined as $V_\text{HOM}=1-2g^{(2)}_{\text{HOM},\parallel}$, which implies $M=V_\text{HOM}+g^{(2)}$. Both approaches should predict identical values of $M$. Furthermore, if $g^{(2)}$ is small but nonzero and one of the two photons is approximately distinguishable from the emitted single photon state \cite{ollivier2021hong}, then $M$ is related to $M_\mathrm{s}$ via $M \simeq M_\mathrm{s} (1-g^{(2)})$. Hence, in our experiments, $M_\mathrm{s}$ is accurately estimated by
\begin{equation}
    M_\mathrm{s} \simeq V_\text{HOM}\frac{1+g^{(2)}}{1-g^{(2)}}.
\end{equation}}

\begin{figure}
	\centering
	\includegraphics[width=0.9\linewidth]{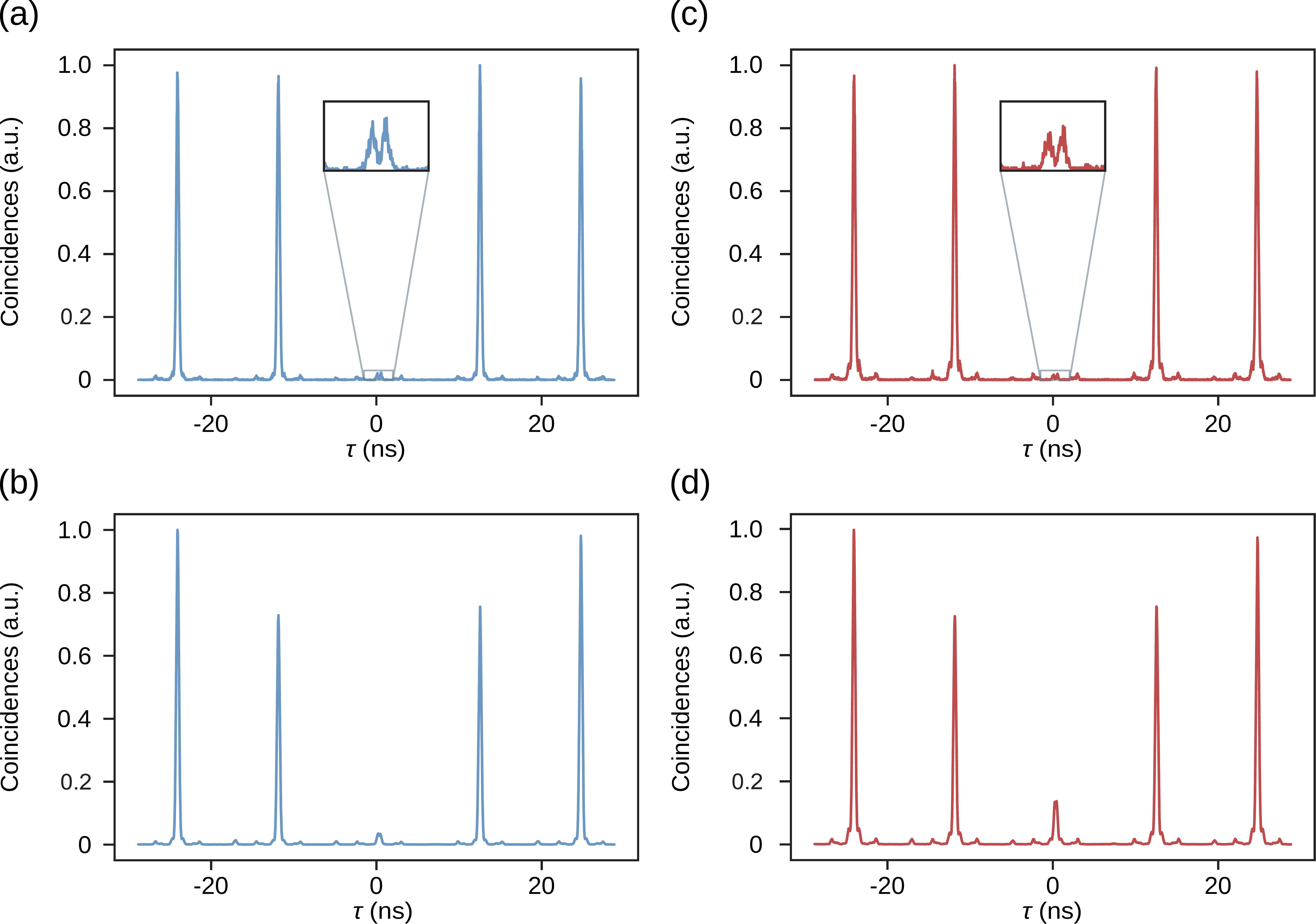}
	\caption{\textbf{ Performance of Single-Photon Source}  (a) Auto-correlation measurement of the single-photon wave packets generated by the single-photon source with $\theta = \pi$ at 5~K. The resulting coincidence histogram allows for extraction of the single-photon purity $g^{(2)}$. (b) The coincidence histogram obtained from Hong-Ou-Mandel interference measurements $g^{(2)}_\mathrm{HOM}$ for $\theta = \pi$ at 5~K. From Eq. 24 and the single-photon purity extracted from (a) we can deduce the mean wave packet overlap, {or single-photon} indistinguishability, $M_\mathrm{s}$. (c) and (d) the same as (a) and (b) but at 20~K. {Note that the small peaks observed in the coincidence histograms at non-integer multiples of the pulse separation time of 12.5~ns are due to electronic reflections in our measurement setup and can be neglected.}\label{fig:S1}}
	\end{figure}

Fig.~\ref{fig:S1} shows the results of these {second-order intensity  correlation} measurements with Fig.~\ref{fig:S1}(a) and Fig.~\ref{fig:S1}(b) (Fig.~\ref{fig:S1}(c) and Fig.~\ref{fig:S1}(d)) { corresponding to the auto-correlation} $g^{(2)}$, and {the time integrated HOM measurement} $g^{(2)}_\mathrm{HOM}$ at 5~K (20~K), all taken at $\theta=\pi$. We extract similar auto-correlation values at 5~K and 20~K: $g^{(2)}$ of $(2.84\pm0.08)\%$ and $(2.28\pm0.08)\%$, respectively. However, the indistinguishability between two subsequently emitted single-photon wave packets is  reduced at the higher temperature. From the indistinguishability measurements we deduce a {single-photon indistinguishability} of $M_\mathrm{s} = (92.6 \pm 0.1)\%$ at 5~K and $M_\mathrm{s} = (58.0 \pm 1.0)\%$ at 20~K.

\subsubsection*{IV.3. Photonic and classical fields overlap}

\begin{figure}
		\centering
		\includegraphics[width=0.9\linewidth]{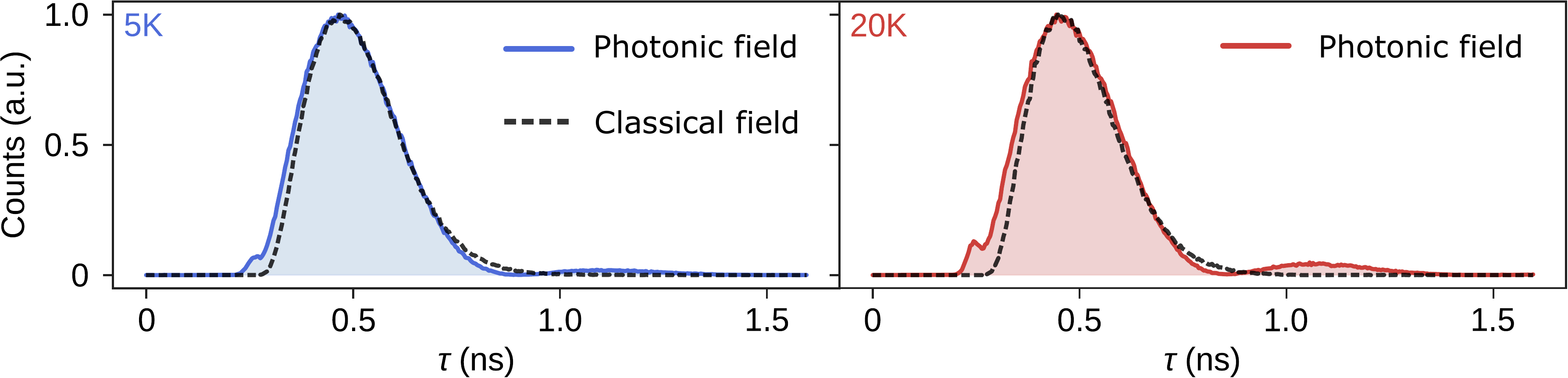}\caption{\textbf{Temporal profiles photonic field and classical field.} The classical field (dashed black), derived from the same coherent laser field used to drive the quantum dot, is temporally shaped to overlap with the temporal profile of the emitted photonic field at 5K and 20K (blue and red, solid).}\label{fig:temporalprofile}	
\end{figure}

{To generate temporal mode overlap between the emitted photonic field and the classical field we shape the latter with a Fabry-P\'erot etalon. By time resolving the emission dynamics of both fields (see Fig.~\ref{fig:temporalprofile}), we qualitatively ensure temporal overlap between the classical field (black dashed) and the photonic field (blue 5~K, red 20~K).}
{To then quantify the mean wave packet overlap between the two fields, $M_\text{f,c}$, we perform a HOM experiment with the two fields for $\theta=\pi$. Here it is} important to account for the non-negligible classical intensity correlation $g^{(2)}_\mathrm{c}=1$ of the coherent state input. The coincidence counts can arise in three ways. (1) A photon from each input do not bunch when leaving the beam splitter, leading to a contribution of $2\mu_\text{f}\mu_\mathrm{c}(1-M_\text{f,c})$. The factor of two in this term arises from the two ways to obtain a coincidence count (both photons reflected or both photons transmitted). (2) Two photons arrive from the coherent state when the photonic field is vacuum (or lost), leading to an additional contribution $\mu_\mathrm{c}^2g^{(2)}_\mathrm{c}=\mu_\mathrm{c}^2$ to the coincidence counts. (3) Two photons arrive from the photonic field input, leading to a small contribution $\mu_\text{f}^2g^{(2)}$ to the coincidence counts. If we now assume that $\theta$ is near $\pi$-pulse, the average uncorrelated histogram peak area $\overline{A}_{\tau>0,\parallel}$ is given by $2\mu_\mathrm{c}\mu_\text{f}$. Thus, for the co-polarized case we have $g^{(2)}_{\mathrm{HOM},\parallel} =1-M_\text{f,c}+\overline{g}^{(2)}$,
where $\overline{g}^{(2)}=(\mu_\text{c}/\mu_\mathrm{f})g^{(2)}_\text{c}/2+(\mu_\mathrm{f}/\mu_\text{c})g^{(2)}_\mathrm{f}/2$ is the weighted average input intensity correlation. The intensity correlation after interference of inputs in cross-polarization is then simply obtained by $g^{(2)}_{\mathrm{HOM},\perp} =1+\overline{g}^{(2)}$. Hence, $M_\text{f,c}$ is measured by
\begin{equation}
    M_\text{f,c}=\frac{g^{(2)}_{\mathrm{HOM},\perp}-g^{(2)}_{\mathrm{HOM},\parallel}}{g^{(2)}_{\mathrm{HOM},\perp}}\left(1+\overline{g}^{(2)}\right).
    \label{eq:wvpMfc}
\end{equation}
When the inputs are balanced ($\mu_\mathrm{c}=\mu_\text{f}$), we have $1+\overline{g}^{(2)}=1+(g^{(2)}+1)/2$. If the photonic field's $g^{(2)}\ll 1$, the correction factor becomes 3/2.

\begin{figure}
		\centering
		\includegraphics[width=0.9\linewidth]{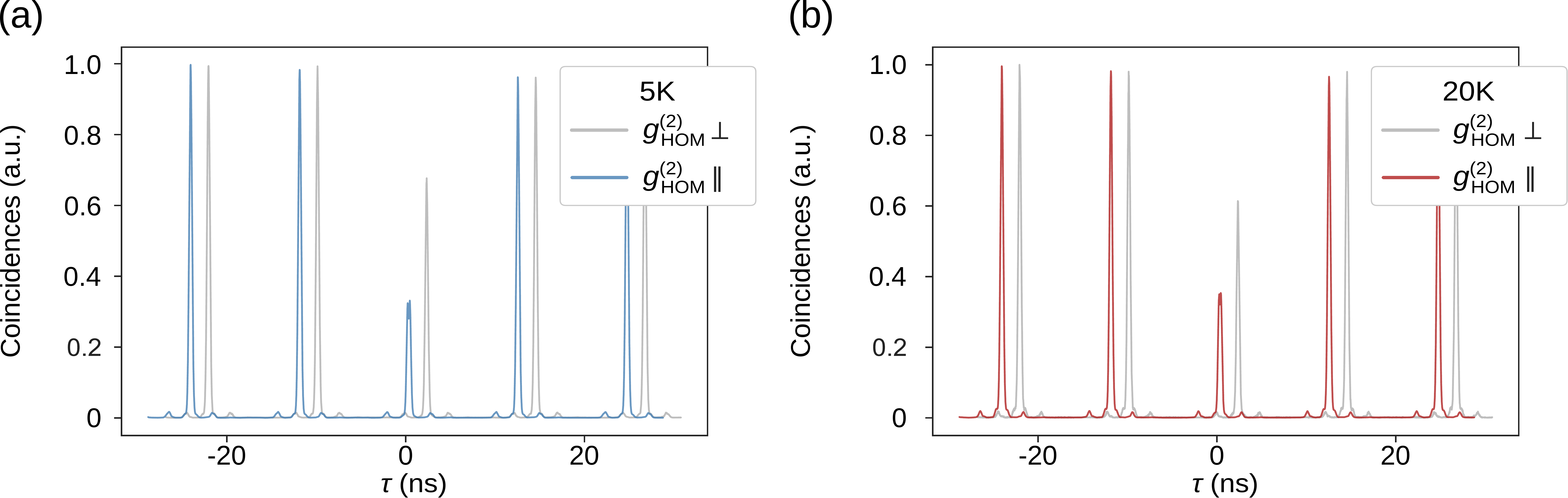}	
		\caption{\textbf{ Mean wave packet overlap between classical coherent field and single-photon field.} (a) The coincidence histograms of time integrated Hong-Ou-Mandel interference measurements between the single-photon field (generated with pulse area $\theta=\pi$) and the classical coherent field at 5~K, measured in co- ($g^{(2)}_\mathrm{HOM} \parallel$) and cross- ($g^{(2)}_\mathrm{HOM} \perp$) polarization configuration. (b) Same as (a) but at 20~K. Using Eq.~\ref{eq:wvpMfc} we obtain mean wave packet overlaps of $M_\text{f,c} = (48.9 \pm 0.3)\%$ at 5~K and $M_\text{f,c}= (32.3 \pm 0.7)\%$ at 20~K. \label{fig:S2}}
\end{figure}

Fig.~\ref{fig:S2} shows the coincidence histograms of HOM measurements with the quantum photonic field and the classical field at 5~K and 20~K, Fig.~\ref{fig:S2}(a) and Fig.~\ref{fig:S2}(b), respectively. We perform the measurements with the two fields cross-polarized ($g^{(2)}_{\mathrm{HOM},\perp}$, grey), and co-polarized ($g^{(2)}_{\mathrm{HOM},\parallel}$, blue and red). From the coincidence histograms we extract a mean wave packet overlap of $M_\text{f,c} = (48.9 \pm 0.3)\%$ at 5~K and $M_\text{f,c}= (32.3 \pm 0.7)\%$ at 20~K. 

\subsubsection*{IV.4. Measuring $\mathcal{C}$ and $\mathcal{C}_\mathrm{f,c}$}

\begin{figure}
		\centering
		\includegraphics[width=0.9\linewidth]{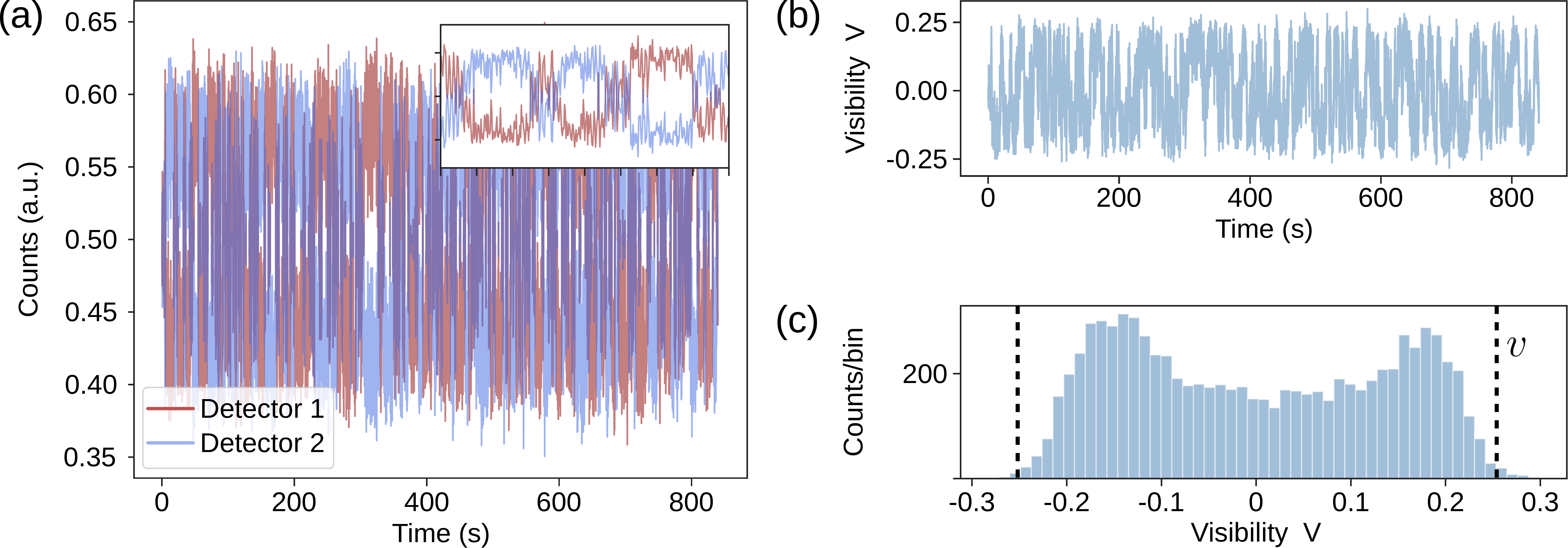}	
		\caption{\textbf{ Extracting visibility of interference $v$}. (a) Intensities measured by detector 1 (red) and 2 (blue) {at the output of the interference beam splitter (BS1 in Figure 1b)} as a function of time, where each point at $t_i$ is normalized by the sum of the two intensities at $t_i$. The inset shows a zoom of the signal, demonstrating anti-correlated intensities. Over the full measurement time we let the relative phase between the two signals freely evolve. (b) The visibility $V$ as a function of time obtained from Eq.~\ref{eq:VSHD} corresponding to the data shown in (a). (c) Histogram of visibility time trace presented in (b). Dashed vertical lines indicate the extracted maximum visibility $v$. This data set was taken at $\theta =0.55\pi$. \label{fig:S3}}
\end{figure}

To determine both $\mathcal{C}$ and $\mathcal{C}_{\mathrm{f,c}}$, homodyne measurements are performed where we measure the difference in count rate at each detector as the interferometer phase evolves. We detail here the protocol for the self-homodyne measurement (step 1). The same protocol applies for the {energetic transfers in spontaneous emission} with some small adjustments as indicated later on.
The maximum visibility of these interference fringes $v$ give the integrated first-order coherence $c^{(1)}$~\cite{Loredo2019,Wein2021PhotonnumberEG} of the photonic field, where $v\equiv  c^{(1)}= (1/\mu_\mathrm{f})\int dt\left|\braket{\hat{a}_\mathrm{f}(t)}\right|^2$. If the mode contains no more than one photon, then $\braket{\hat{a}_\mathrm{f}(t)}=\sqrt{p_0p_1}\zeta(t)$ and $\mu_\mathrm{f}=p_1$. Thus, $\mathcal{C}$ can be determined from $c^{(1)}=p_0\mathcal{C}$.

Experimentally, we interfere two fields which are matched in intensity, polarization, and time-of-arrival at a 50:50 beam splitter. We monitor the intensity at the two outputs of the beam splitter using two superconducting nanowire single photon detectors (SNSPDs, Single Quantum Eos), which register the single-photon counts as a function of time, with 100~ms resolution. When interfering two fields {that show photon-number coherence}, the count rates at the two outputs of the beam splitter are found be to anti-correlated. If the two fields are in (out of) phase then we see maximum constructive (destructive) interference in output 1, corresponding to a maximum (minimum) count rate in detector 1, respectively. In order to determine the maximum interference visibility we record the intensities measured by the two detectors for approximately 20 minutes whilst we let the phase between the two input fields, $\phi$, freely evolve, to ensure that the full phase space has been explored. An example of the intensities measured by two detectors as a function of time for $\theta=0.55\pi$ is given in Fig.~\ref{fig:S3}(a) with red (blue) being the intensity measured by detector 1 (2). The inset of Fig.~\ref{fig:S3}(a) shows a zoom of the signal, clearly displaying anti-correlated intensity signals. The measured intensities vary over time due to the freely evolving phase.

Fig.~\ref{fig:S3}(b) shows the extracted visibility (\verb+Eq. 9+) corresponding to the raw data shown in Fig.~\ref{fig:S3}(a). We build a histogram of the measured visibility values over time, as shown in Fig.~\ref{fig:S3}(c). We extract the maximum visibility of interference, $v$, by taking the average of the absolute value of the $N$ highest, and $N$ lowest visibility data points (with $N=100$). The error is calculated through error propagation and standard counting error. For the example data set shown in Fig.~\ref{fig:S3}, we have indicated in Fig.~\ref{fig:S3}(c) the corresponding $v$ with two vertical dashed lines and label. \\

If the wave packet has a small probability of containing more than one photon, then $v\equiv  c^{(1)}$ can overestimate $\mathcal{C}$. To correct for this effect, we use the approach detailed in Ref.~\cite{ollivier2021hong} and consider that a small amount of noise is added to the ideal state in Eq.~\ref{generalDensity} by a beam splitter interaction. We then decompose the photonic field amplitude $\braket{\hat{a}_\mathrm{f}(t)}$ into a contribution from the desired quantum state (subscript $\mathrm{s}$) and the additional noise (subscript $\mathrm{n}$) $\braket{\hat{a}_\mathrm{f}(t)} = \cos(\vartheta)\braket{\hat{a}_\text{s}(t)}+\sin(\vartheta)\braket{\hat{a}_\text{n}(t)}$, where $\vartheta$ is a noise parameter governing the amount of added noise. Then, the integrated first-order coherence becomes
\begin{equation}
    \mu_\mathrm{f}c^{(1)} =p_1p_0\mathcal{C}\cos^2(\vartheta) +2\sqrt{p_1\mu_\text{n}}c^{(1)}_{\text{s,n}}\cos(\vartheta)\sin(\vartheta)+\mu_\text{n}c^{(1)}_\text{n}\sin^2(\vartheta),
\end{equation}
where $\mu_\mathrm{f}=\mu_\mathrm{s}+\mu_\mathrm{n}=p_1+\mu_\mathrm{n}$, $c^{(1)}_\text{s,n}=(1/\sqrt{\mu_\mathrm{s}\mu_\mathrm{n}})\int dt\text{Re}\left[\braket{\hat{a}_\mathrm{s}(t)}\braket{\hat{a}_\mathrm{n}(t)}\right]$ quantifies the first-order coherence between the noise and the quantum state, and $c^{(1)}_\text{n}=(1/\mu_\mathrm{n})\int dt\left|\braket{\hat{a}_\mathrm{n}(t)}\right|^2$ quantifies the first-order coherence of the classical noise itself. In our experiments, the noise arises from reflected unfiltered laser from the fast qubit excitation pulse, which is temporally separate from the light emitted into the mode of the electromagnetic field by the qubit. Hence, the noise is not coherent with the quantum state $c^{(1)}_\mathrm{s,n}=0$ but is itself coherent by definition $c^{(1)}_\mathrm{n}=1$. In this notation,  $g^{(2)}\ll 1$ can be similarly written as in the supplementary of Ref.~\cite{ollivier2021hong}
\begin{equation}
\begin{aligned}
    \mu_\mathrm{f}^2g^{(2)} = 2(1+M_\text{s,n})p_1\mu_\text{n}\cos^2(\vartheta)\sin^2(\vartheta)+\mu_\mathrm{n}^2\sin^4(\vartheta)
\end{aligned}
\end{equation}
where $M_\text{s,n}=(1/\mu_\mathrm{s}\mu_\mathrm{n})\iint dtdt^\prime\text{Re}\left[G^{(1)}_\mathrm{s}(t,t^\prime)G^{(1)}_\mathrm{n}(t^\prime,t)\right]\simeq 0$ is the mean wave packet overlap between the emitted quantum state and the temporally separate classical noise photons. Note that here we used $g^{(2)}_\mathrm{s}=0$ and $g^{(2)}_\mathrm{n}=1$. Defining $\cos^2(\eta)=p_1\cos^2(\vartheta)/\mu_\mathrm{f}$ and $\sin^2(\eta)=\mu_\text{n}\sin^2(\vartheta)/\mu_\mathrm{f}$, we can re-write our expressions in terms of a single parameter:
\begin{equation}
\begin{aligned}
    c^{(1)}(\eta)&= p_0\mathcal{C}\cos^2(\eta) +\sin^2(\eta)\\
    g^{(2)}(\eta)&= 2\cos^2(\eta)\sin^2(\eta)+\sin^4(\eta)
\end{aligned}
\end{equation}
Clearly, we have $c^{(1)}=p_0\mathcal{C}$ when $\eta\rightarrow 0$ as expected for the ideal case. If nonzero, the lowest-order correction is then given by $\lim_{\eta\rightarrow 0}(dc^{(1)}(\eta)/dg^{(2)}(\eta)) = (1-p_0\mathcal{C})/2$.
Hence we have

\begin{equation}
    p_0\mathcal{C}\simeq\frac{c^{(1)}-g^{(2)}/2}{1-g^{(2)}/2}.
\end{equation}

\noindent Note that this formalism allows to account for the residual $g^{(2)}$ in the visibility measurement (blue and red curves in  Fig.~\ref{fig:2}(a) and Fig.~\ref{fig:2}(b)).{ Indeed, a small fraction of laser signal in the measured field leads to classical interference that artificially increases the amount of unitary work transferred. This effect contributes all the more as we increase $\theta$ as attested by an increased second-order intensity correlation of the emitted photonic field. }

\newpage
\subsection*{V. Glossary}

\renewcommand{\arraystretch}{1.5}
\begin{table}[h]
\caption{List of terms, symbols, and definitions for quantities related to energetics.}\footnotesize
\begin{tabular}{l|c|l}
     Name & Symbol & Definition or constraint\\\hline\hline
     Qubit initial state & $\ket{\Psi_\mathrm{q}}$ & $\cos(\theta/2)\ket{g}+\sin(\theta/2)e^{i\phi}\ket{e}$\\
     Qubit initial energy & $\mathcal{E}_\mathrm{q}$ & 
     $\hbar\omega_0\sin^2(\theta/2)$\\
     Qubit initial coherence & $s$ & $\cos(\theta/2)\sin(\theta/2)$\\
     Photonic field mode & $\hat{a}_\mathrm{f}$ & -\\
     Photonic field envelope (intensity profile) & $I_\mathrm{f}(t)$ & $\braket{\hat{a}^\dagger_\mathrm{f}(t)\hat{a}_\mathrm{f}(t)}$\\
     Photonic field initial energy & $\mathcal{E}_\mathrm{f}$ & $\hbar\omega_0 \mu_\mathrm{f}=\mathcal{E}_\mathrm{q}(0)$\\
     Photonic field ideal state & $\ket{i_\mathrm{f}}$ & $\cos(\theta/2)\ket{0}+\sin(\theta/2)e^{i\phi}\ket{1}$ \\
     Photonic field ideal unitary energy &$E^{\mathrm{q,f}}_\mathrm{unit}$ & $\hbar\omega_0 s^2$\\
     Qubit-Photonic field energy transfer efficiency & $\eta_\mathrm{q,f}$ & $E^\mathrm{q,f}_\mathrm{unit}/\mathcal{E}_\mathrm{f}$ \\
     Classical field mode (coherent state) & $\hat{a}_\mathrm{c}$ & -\\
     Classical field amplitude & $\beta(t)$ & $\braket{\hat{a}_\mathrm{c}(t)}$\\
     Classical field envelope (intensity profile) & $I_\mathrm{c}(t)$ & $\braket{\hat{a}_\mathrm{c}^\dagger(t)\hat{a}_\mathrm{c}(t)}=|\beta(t)|^2$\\
     Classical field initial energy & $\mathcal{E}_\mathrm{c}$ & $\hbar\omega_0 \mu_\mathrm{c}$\\
     \hline\hline
\end{tabular}
\end{table}

\begin{table}
\caption{List of terms, symbols, and definitions for quantities for the analysis of photonic states.}\footnotesize
\begin{tabular}{l|c|l}
     Name & Symbol & Definition or constraint\\\hline\hline
     Photon creation operator (time basis)& $\hat{a}^\dagger(t)$ & $[\hat{a}(t),\hat{a}^\dagger(t^\prime)]=\delta(t-t^\prime)$\\
     Photonic density operator (number basis) & $\hat{{\rho}}$ & $p_0\hat{{\rho}}_0 + p_1\hat{{\rho}}_1 + \sqrt{p_0p_1}(\hat{{\rho}}_{01}+\hat{{\rho}}_{10})+\cdots$\\
     Vacuum state & $\hat{{\rho}}_0$ & $\ket{0}\!\bra{0}$\\
     Single-photon state & $\hat{{\rho}}_1$ & $\iint dt dt^\prime\xi(t,t^\prime)\hat{a}^\dagger(t)\ket{0}\!\bra{0}\hat{a}(t^\prime)$\\
     Single-photon temporal density function & $\xi(t,t^\prime)$ & 
     $\text{Tr}[\hat{a}^\dagger(t^\prime)\hat{a}(t)\hat{\rho}_1]$\\
     Number coherence (between $\ket{0}$ and $\ket{1}$) & $\hat{{\rho}}_{01}$ & $\int dt \zeta(t)\hat{a}^\dagger(t)\ket{0}\!\bra{0}$\\
     Temporal number coherence amplitude & $\zeta(t)$ & 
     $\text{Tr}\left[\hat{a}(t)\hat{\rho}_{01}\right]$
     \\
     Temporal wave packet (first-order correlation)$^*$ & $G^{(1)}(t,t^\prime)$ &
     $\braket{\hat{a}^\dagger(t^\prime)\hat{a}(t)}$\\
     Temporal envelope (intensity profile) & $I(t)$ & $\braket{\hat{a}^\dagger(t)\hat{a}(t)}$\\
     Average photon number & $\mu$ 
     & $\int dtI(t)=\sum_nnp_n$ \\
     Mean wave packet overlap & $M_{i,j}$ & $\frac{1}{\mu_i\mu_j}\iint dtdt^\prime\text{Re}\left[G^{(1)}_i(t,t^\prime)G^{(1)}_j(t^\prime,t)\right]$\\
     Indistinguishability & $M$ & $\frac{1}{\mu^2}\iint dtdt^\prime\left|G^{(1)}(t,t^\prime)\right|^2$\\
     Single-photon indistinguishability (trace purity) & $M_\mathrm{s}$ & $\iint dtdt^\prime \left|\xi(t,t^\prime)\right|^2=\text{Tr}[\hat{{\rho}}_1^2]$\\
     Normalized$^\dagger$ first-order coherence amplitude overlap& $c^{(1)}_{i,j}$ & $\frac{2}{\mu_i+\mu_j}\int dt\text{Re}\left[\braket{\hat{a}_i(t)}\braket{\hat{a}_j^\dagger(t)}\right]$\\
     Number purity & $\mathcal{C}$ & $\int dt \left|\zeta(t)\right|^2=\text{Tr}\left[\hat{{\rho}}_{01}\hat{{\rho}}_{10}\right]$, $0\leq \mathcal{C}\leq \sqrt{M_\mathrm{s}}$\\
     Number purity parameter & $\lambda^2$ & $\mathcal{C}/\sqrt{M_\mathrm{s}}$, $0\leq \lambda \leq 1$\\
     Second-order (intensity) correlation & $G^{(2)}(t,t^\prime)$ & $\braket{\hat{a}^\dagger(t)\hat{a}^\dagger(t^\prime)\hat{a}(t^\prime)\hat{a}(t)}$\\
     Normalized time-integrated second-order correlation$^\ddagger$ & $g^{(2)}(\tau)$ & $\frac{1}{\mu}\int dt G^{(2)}(t,t+\tau)$\\
     Normalized fully integrated second-order correlation$^\ddagger$ & $g^{(2)}$ & $\frac{1}{\mu^2}\iint dtdt^\prime G^{(2)}(t,t^\prime)=\frac{2}{\mu^2}\sum_n \binom{n}{2}p_n$\\
     \hline\hline
\end{tabular}\\
$^*$ Upper-case $G$ is used for unnormalized correlations, lower-case $g$ is used for normalized correlations.\\
$^\dagger$ An alternative convenient normalization is $1/\sqrt{\mu_i\mu_j}$, which is equivalent when $\mu_i=\mu_j$.\\
$^\ddagger$ Note that $g^{(2)}\neq g^{(2)}(0)$ in this notation.
\end{table}

\end{document}